\documentclass[preprint,12pt,authoryear]{elsarticle}

\usepackage{epsfig}

\usepackage{wasysym}

\usepackage{amssymb}

\usepackage{natbib}

\usepackage{color}
\usepackage{lscape}

\newcommand{\vunit}{\mbox{\,km\,s$^{-1}$}}

\newcommand{\Msun}{\mbox{\,M$_\odot$}}

\newcommand{\mic}{\mbox{$\,\mu$m}} 
\newcommand{\chemone}{\raisebox{0.03cm}{$-$}} 

\newcommand{\ltsimeq}{\raisebox{-0.6ex}{$\,\stackrel 
{\raisebox{-.2ex}{$\textstyle <$}}{\sim}\,$}} 
\newcommand{\gtsimeq}{\raisebox{-0.6ex}{$\,\stackrel
{\raisebox{-.2ex}{$\textstyle >$}}{\sim}\,$}}

\newcommand{\fion}[2]{[{#1}\,{\sc {#2}}]}

\newcommand{\nucl}[2]{\mbox{$^{#1}$}{#2}}

\DeclareGraphicsExtensions{.png,PNG,.pdf,.PDF,.ps}

\begin{document}

\begin{frontmatter}

\title{Pre-solar grains from novae and ``Born-again giants''}


\author[ku]{A. Evans}

\author[mn]{R. D. Gehrz}

\address[ku]{Astrophysics Group, Lennard-Jones Laboratories, Keele University, 
Keele, Staffordshire, ST5 5BG, UK}

\address[mn]{Minnesota Institute for Astrophysics
John T. Tate Hall 116 Church St. SE Minneapolis, MN 55455, USA}

\begin{abstract}
We review the properties of dust formed during classical nova
eruptions and the Very Late Thermal Pulses (VLTPs) that occur 
during the later stages of post-Asymptotic Giant Branch 
evolution of low-mass stars. In both cases, carbon and 
hydrocarbon dust is produced. Novae may also 
produce silicate dust, contrary to the usual paradigm about 
the C:O ratio and dust composition. Despite the expectation that
these dust sources are not expected to make significant 
contributions to the Galactic dust population, there is a 
significant body of evidence that grains from both stellar 
sources have been identified in recovered 
meteoritic and cometary material, and that certain 
infrared spectral signatures 
seen in comets are common to novae, VLTPs and 
pre-solar grains.
 \end{abstract}


\end{frontmatter}

\tableofcontents
%

 \section{Introduction}
Nova eruptions and Very Late Thermal Pulses (VLTPs) in low mass stars
originate on the surfaces of degenerate objects: white dwarfs (WDs) 
in binary systems in novae, and on top of the degenerate cores of 
Asymptotic Giant Branch (AGB) stars in VLTPs. In novae and VLTPs 
the nuclear burning leads to an eruption,
which is explosive in the case of novae. When VLTPs are
reported, it is common for them to be described as a ``nova''. For example
the VLTP V4334~Sgr\footnote{In general, variable star names in this chapter
follow the convention recommended by the International Astronomical Union.}
(known as Sakurai's Object, in honour of its discoverer,
the amateur astronomer Yukio Sakurai) was, until 2021, still listed in the SIMBAD\footnote{http://simbad.u-strasbg.fr/simbad/} 
database as a ``nova-like star''; at the time of this writing
it is listed as a ``Cataclysmic Variable Star'', which is 
somewhat less inaccurate.

Evidence in Solar System material for ``stardust'', grains produced in 
stellar winds, comes from (a)~laboratory studies 
of recovered material (such as meteorites and 
{interplanetary dust particles, hereafter IDPs)}, 
in which elemental and isotopic signatures are identified and 
(b)~spectroscopic observations, comparing (especially) infrared (IR)
spectra of Solar System objects and extra-solar proto-planetary
systems with the IR spectra of the sources of stardust. 
Regarding elemental and isotopic abundances, it is
preferable to have agreement 
between predicted and measured values for as many species as possible
in order to discriminate between stellar sources. For example, the
\nucl{12}{C}/\nucl{13}{C} ratio for both novae and VLTPs is predicted
to be $\ltsimeq10$ (see below). 

In this chapter we briefly describe the causes of nova and VLTP
eruptions, the way in which the eruption progresses, the nature of 
the ejected material and of the dust formed. In each case we 
describe the evidence indicating that these systems contributed to 
the material from which the Sun and its entourage formed.

Given the considerable history of observation and theory for novae
by comparison with VLTPs, our discussion is heavily biased towards 
novae.

\section{Nova eruptions}

\subsection{The eruption\label{eruption}}

Classical nova eruptions occur in semi-detached binary systems 
consisting of a WD primary and a late-type secondary 
\citep[see comprehensive reviews by][]{CN2,BASI,woudt14}. The WD 
may be of ``CO'' or of ``ONe'' type, the latter generally, but not 
exclusively, being at the high-mass
end of WD masses. The late-type star is usually a 
main sequence dwarf, but nova 
systems are known that have sub-giant secondaries 
\citep[e.g., GK~Per;][]{kraft64}. 

The secondary star fills its Roche lobe;
material from the secondary flows onto the surface of the WD through
the inner Lagrangian point, and spirals onto the WD surface as an 
accretion disk. In time the material at the base of the accreted 
layer becomes degenerate, and hot enough to initiate hydrogen burning. 
A Thermonuclear Runaway (TNR) ensues, resulting in the 
explosive ejection of some $10^{-6}$--$10^{-4}$\Msun\ of material, 
at several 100 to several 1000\vunit. As a result of the TNR, and the
ingestion of some WD material into the TNR region, the ejected 
material is enriched in C, N, O, Mg, Si, Al, Ne, and other metals. 
See Jordi Jos\'e's contribution to this volume 
(Chapter~9), and references therein, for a full discussion of the TNR.

The ejecta, both as gas and dust that condenses as the 
expanding ejecta cool, enrich the interstellar medium. Moreover, 
the TNR imprints nova-specific isotopic ratios on the ejecta, and 
it is likely that classical nova explosions are a major source of 
Galactic \nucl{13}{C}, \nucl{15}{N} and \nucl{17}{O} \citep[see][]{gehrz98}.  
It has also been suggested that novae
might make a significant contribution to Galactic \nucl{7}{Li}
\citep[][and references therein]{starrfield20}, although the case is far
from clear \citep[][and Chapter~9]{jose20}.

In time (typically $\sim5-10$~years), the eruption subsides and 
mass-transfer resumes. The accretion disk is re-established until 
the conditions again arise for another TNR to occur, after an interval of 
$\sim10^4-10^5$~years. All novae are therefore recurrent, but where 
the secondary is a red giant the recurrence time-scale is less 
than a human lifetime, so these systems are seen by humans to 
erupt more than once; these are the ``recurrent novae'' 
\citep[RNe; see][]{RSO}. In general, these are not dust-formers, although 
there was an extremely transient phase of CO and dust formation
during the 2014 eruption of the RN V745~Sco \citep{banerjee23}. The
recurrent nova RS~Oph displays strong 9.7\mic\ and 18\mic\ silicate
features, which are associated with the red giant secondary 
\citep{evans07}. Henceforth we largely confine our discussion to 
classical novae.

The mass-loss from a nova eruption is in the form of a continuous
but diminshing
wind that originates on the WD surface \citep{bath76}.
A key property of classical nova eruptions is that, to a good 
approximation, the {\em bolometric} luminosity $L_{\rm bol}$ of the 
stellar remnant remains $\sim$~constant throughout the eruption
\citep[as was first shown theoretically by][]{bath76};
this constant bolometric luminosity is of the order of,
or may exceed, the Eddington luminosity 
for the WD. Thus as the pseudo-photosphere of the ejecta shrinks 
back onto the surface of the WD as the mass-loss rate diminishes, 
the effective temperature increases to maintain constant $L_{\rm bol}$. 
The well-known visual light decline of an erupting nova
is therefore simply a bolometric correction effect as the maximum
of the spectral energy distribution (SED), which is roughly
Planckian, shifts out of the visible, into the ultra-violet
and eventually X-rays. The visual magnitude falls as 
the Rayleigh-Jeans tail slides through the visible window 
toward short wavelengths \citep{bath76,gallagher76}. 

\subsection{Abundances in the ejecta\label{nova-abun}}

The composition of the ejecta from a nova eruption is determined,
amongst other factors, by the composition of the secondary star
(material from the surface of which is deposited on the surface 
of the WD), the nature of the WD itself (i.e. whether it is of CO
or ONe type), and how much mixing of the WD and accreted material 
occurs. Moreover, the composition (both elemental and isotopic) 
of the ejecta are important in that they may have infiltrated 
material that produced the proto-solar nebula: 
analysis of material 
{from meteorites and IDPs}
may reveal the extent to which this may have occurred.

\begin{figure*}
\centering
\includegraphics[width=0.85\textwidth]{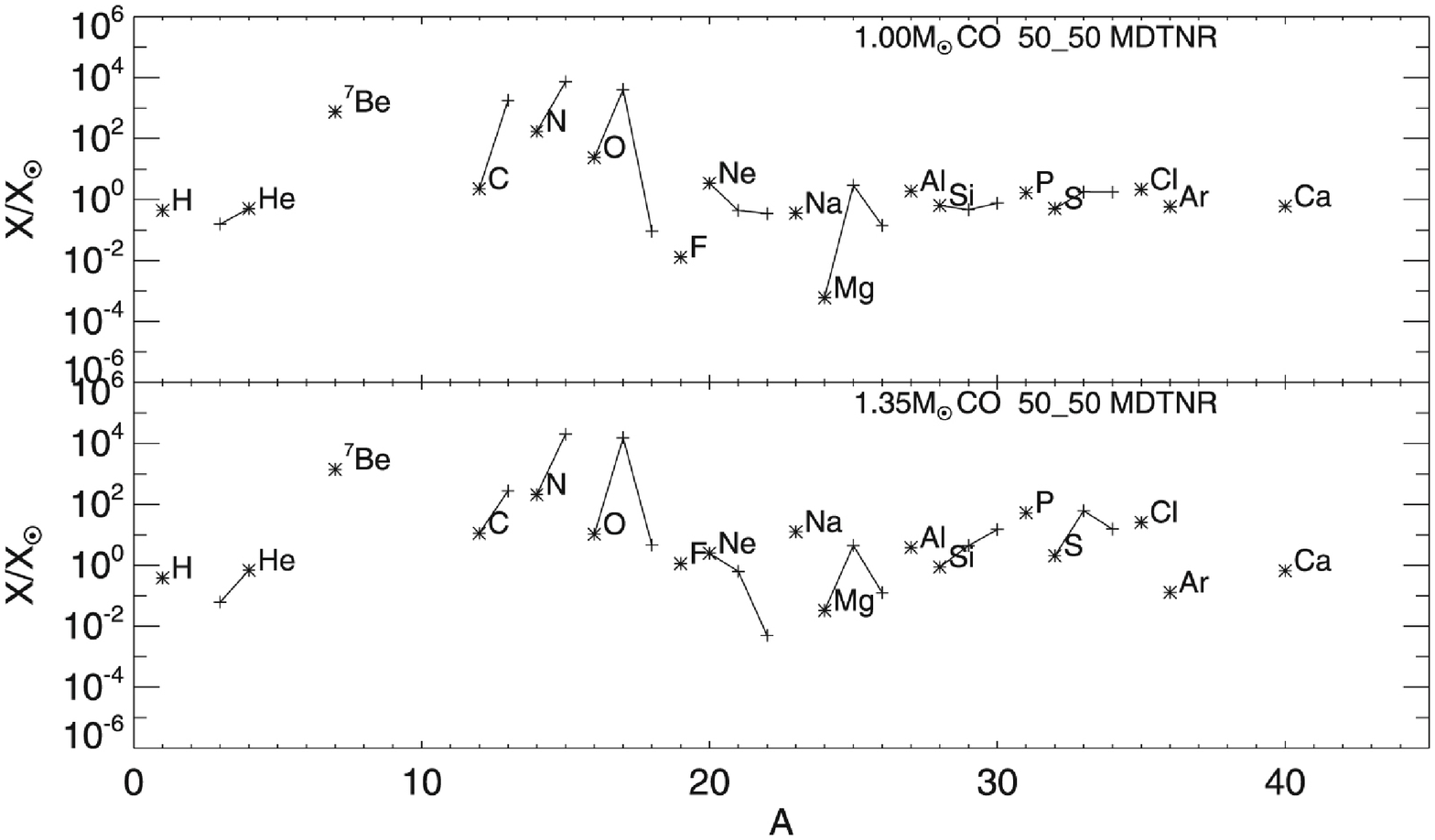}
\includegraphics[width=0.85\textwidth,clip]{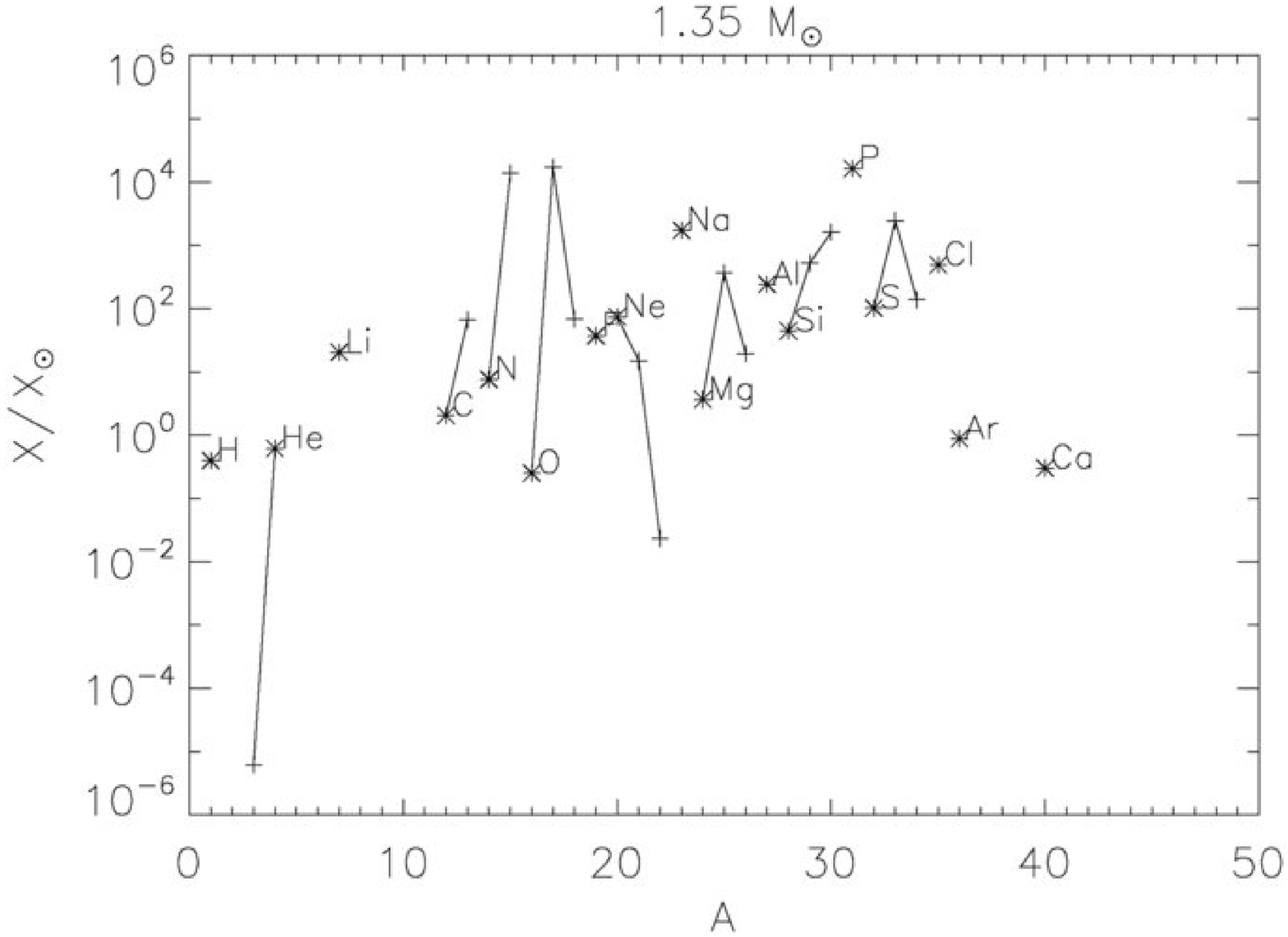}
\caption{Ejecta abundances (by mass) for stable isotopes, from H to Ca,
for a TNR on the surface of a WD in a nova system. $A$ is the atomic
mass; the $y$-axis is the logarithmic ratio of the ejecta abundances
divided by the corresponding solar abundances. 
Top panel: 1.0\Msun\ (top) and 1.35\Msun\ (bottom) CO WD, with a 50:50 mix 
of WD and accreted material. Solar abundances from \cite{lodders03}.
Bottom panel: 1.35\Msun\ ONe WD, with a 50:50 mix of WD and accreted 
material; solar abundances from \cite{anders89}.
The most abundant isotope of a given element is designated by an asterisk; 
all isotopes of a given element are connected by solid lines.
Figures from \cite{starrfield20} (top) and \cite{starrfield09} (bottom).
\copyright{AAS}. Reproduced with permission.
 \label{starrfield}}
\end{figure*}

Simulations for CO and ONe WDs
\citep{starrfield09,starrfield20}, giving the yield in mass
fraction, expressed relative to solar abundances, are shown in 
Fig.~\ref{starrfield}. We note here that care should be taken 
in directly comparing published yields expressed in this way, 
to ensure that the solar benchmarks are the same; for example, 
in Fig.~\ref{starrfield}, the solar reference values are from 
\cite{anders89} (ONe) and \cite{lodders03} (CO) respectively.

\begin{figure*}
\centering
\includegraphics[width=0.95\textwidth]{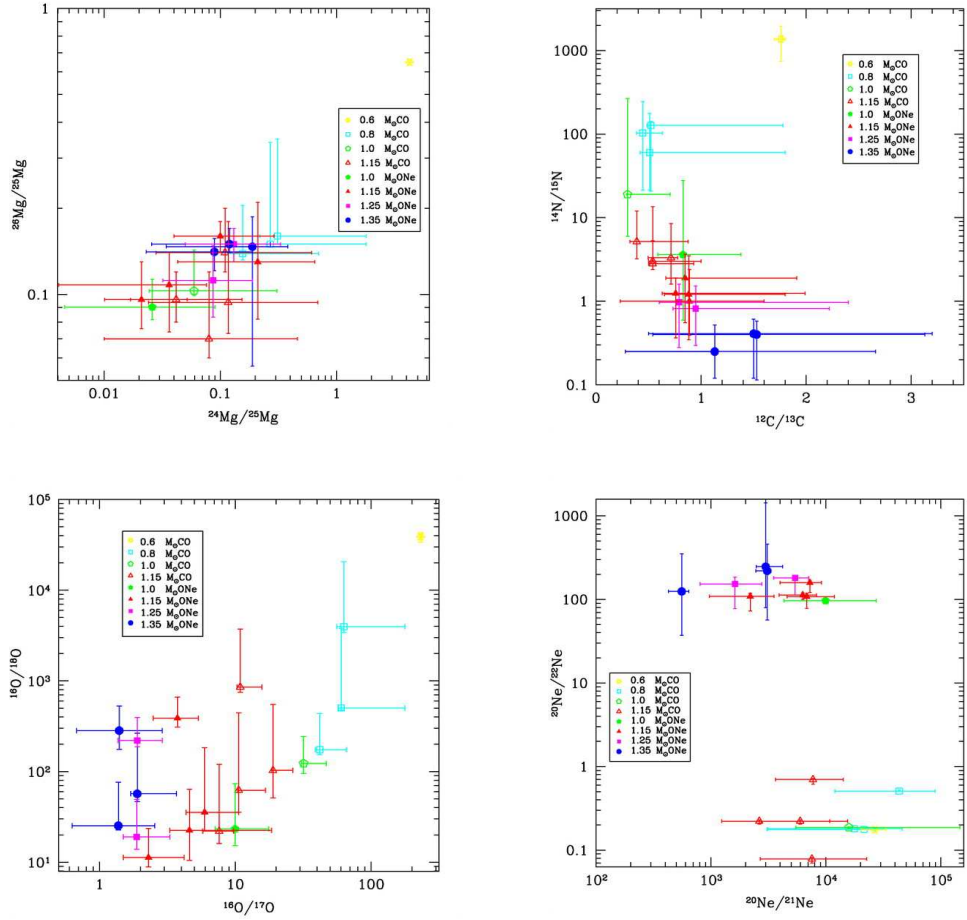}
\caption{Predicted isotopic ratios for TNR for a variety of WD types 
(CO/ONe) and masses. ``Error bars'' take into account the composition
gradient in the ejected shells. From \cite{jose04}.
\copyright{AAS}. Reproduced with permission.
\label{jose_isotopes}}
\end{figure*}

We see from Fig.~\ref{starrfield} that large overabundaces of CNO 
are predicted for CO WDs, suggesting that carbon is a plausible
dust composition for these 
novae\footnote{{Note that
these simulations also show an overabundance of C in ONe novae
(see Fig.~\ref{starrfield}), but ONe novae are generally less
proficient dust producers \citep{gehrz98,evans12}.}}. 
Likewise, large overabundaces of 
Ne, Na, Si, Al, P and S are predicted for ONe WDs. A large overabundance
of Ne was first deduced for nova V1500~Cyg 
\citep{ferland78a,ferland78b}, giving rise to the concept of 
``neon novae'' originating from a TNR on an ONe WD. The implication
is that there is mixing of the WD and accreted material. 
The discovery of large overabundances of Ne in nova ejecta
are now relatively commonplace 
\citep[e.g., QU Vul,][and references therein]{gehrz08}.

Once the ejected material is ionised and excited by the hot stellar
remnant, the nova enters a ``nebular'' phase, in which the 
spectrum at all wavelengths is dominated by strong ionic emission lines
(see e.g., Fig.~\ref{dust_seds} below). At this time, analysis of the 
line fluxes using photoionisation codes such as {\sc cloudy}
\citep{cloudy} allows the determination of elemental abundance relative
to hydrogen. The observations, followed by detailed analysis,
largely confirm the predictions from TNR theory.

Several authors 
\citep[e.g.,][and references therein]{jose04,starrfield09,starrfield20}
have also calculated detailed isotopic ratios for a variety of 
TNRs, on CO and ONe WDs having a range of masses. 
A summary of typical values is shown in Fig.~\ref{jose_isotopes}
\citep[from][]{jose04}. 
These authors find that, for ONe novae, significant overabundances of
\nucl{15}{N} and \nucl{17}{O} are produced, while CO novae may produce
significant amounts of \nucl{7}{Li}
(but see comment in Section~\ref{eruption}).
In the latter novae {\nucl{12}{C}/\nucl{13}{C} is $<1$} for 
CO WD masses in excess of {0.8\Msun}. As pointed out by 
\cite{gehrz08}, a significant amount
of \nucl{22}{Ne} could arise from the radioactive decay of
\nucl{22}{Na} ($\beta^+$ active, half-life 2.60~years) 
produced in the TNR, which is also predicted to be overabundant 
in neon novae.

\subsection{The route to dust formation}

Nova eruptions that occur in systems having CO WDs frequently produce
dust in their ejecta. The evidence for dust formation originally came
from the deep minimum in the visual light curve, first noted in nova 
DQ~Her 1934 \citep[see][]{strope10} in which the dust minimum was 
$\sim10$~mag deep. Around the time of the DQ~Her eruption, 
the deep minima in the light curves of R~Coronae Borealis stars 
\citep[see e.g.,][]{clayton12} was being attributed to obscuration 
by dust \citep{loreta,okeefe}. By 
analogy with the R~CrB stars, it was \cite{mclaughlin35} who first 
suggested that the deep minimum in the visual light curve of DQ~Her 
was due to the formation of dust in the ejected material. However 
it was not until the seminal observations by \cite{geisel70} that 
confirmation of this hypothesis came with the 
detection of a large IR excess in the dust-forming nova 
FH~Ser 1970, that coincided with a deep minimum in its visual 
light curve. Advances in our understanding of the IR temporal 
development of novae have been reviewed by \cite{gehrz98}, \cite{G-CN2},
\cite{banerjee12} and \cite{evans12}.

\begin{figure*}
\centering
 \includegraphics[width=1.0\textwidth,keepaspectratio]{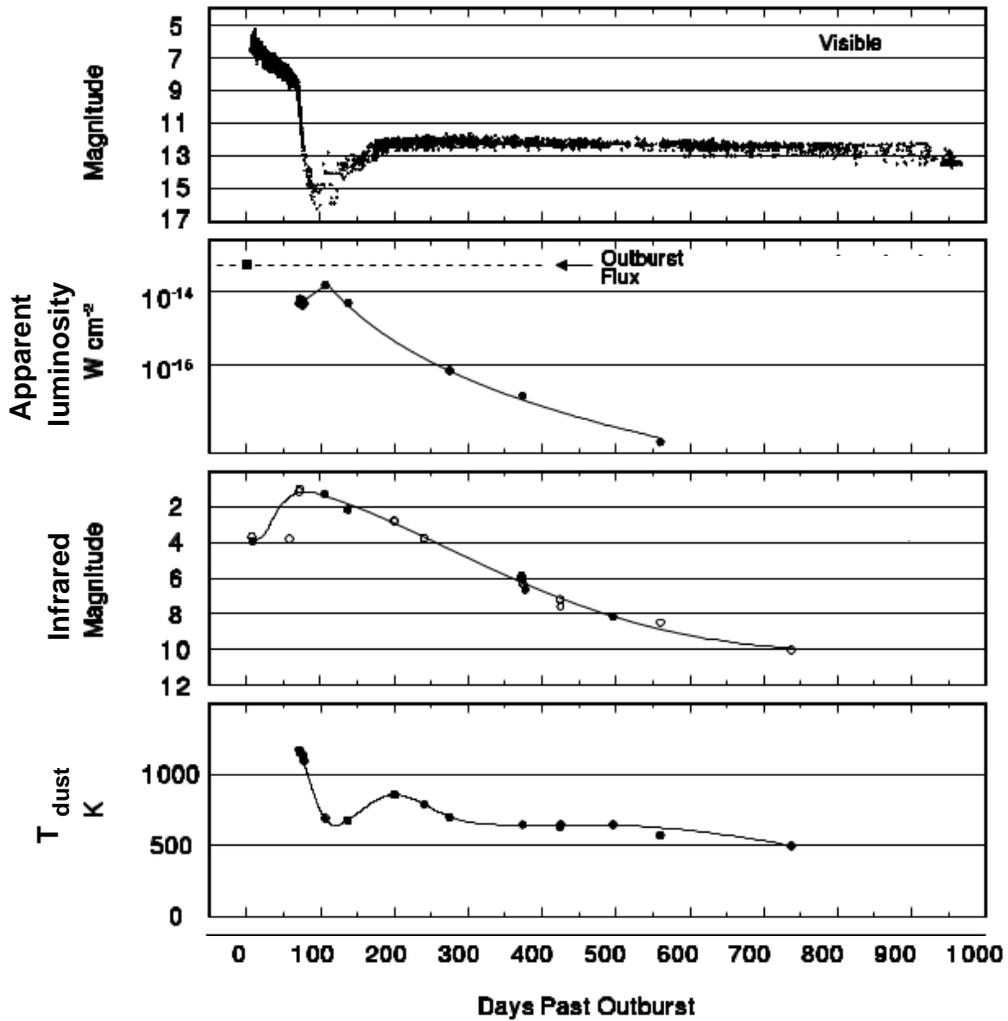}
 \caption{Temporal development of nova V705~Cas. Top and next-to-bottom
 panels illustrate respectively the drop in the visual and the concurrent
 rise in the IR as dust formation occurs. IR magnitudes at 3.6\mic\
 are shown as filled circles; open circles are IR magnitudes at 3.8\mic.
 The dust temperature is given in the bottom panel. 
 Figure adapted from \cite{mason98}. 
 \copyright{AAS}. Reproduced with permission.
\label{nqvul}}
\end{figure*}

There is (as yet) no unbiassed sample of nova light 
curves\footnote{This situation is of course very likely to 
change when the Legacy Survey of Space and Time (LSST) of
the Vera C. Rubin Observatory \citep{lund16} becomes 
operational.}, but the sample in \cite{strope10} probably
comes closest at the time of this writing. The sample compiled 
by \citeauthor{strope10} suggests that the light curves of about 
20\% of classical novae show evidence of deep dust minima, although 
as many as 30\% of classical novae are dust-producers. In some cases, 
the apparent discrepancy likely arises 
becaue the dust distribution is highly anisotropic 
\citep[as in the case of V1280~Sco;][]{chesneau12}: if the optically 
thick part of the dust shell happens not to lie along the line-of-sight 
then only a mild dip in the visual light curve occurs,
even though there may be a substantial IR excess.
In other cases, such as V1668~Cyg \citep{gehrz80b}, a smaller 
amount of dust forms than in the case of novae like V705~Cas
\citep[see Fig.~\ref{nqvul}, adapted from][]{mason98} that 
show deep visual minima during dust formation. 

When the ejecta reach a nominal condensation distance for a specific 
condensate, then dust condensation may occur
if the gas density in the ejected shell is high enough. A necessary 
condition is the formation of nucleation sites and, before that, the
formation of small molecules such as CO and C$_2$. First overtone 
CO emission has frequently been detected 
in erupting novae \citep[see][and references therein]{banerjee16}. 
That paper (their Table~9) also summarises the values of the 
\nucl{12}{C}/\nucl{13}{C} ratio in novae, as determined from the 
CO isotopologue emission. This shows that the ratio is close to unity, 
and far less than the Solar System value of $\sim89$ \citep{asplund09}.
Furthermore, it is evident that, for \nucl{12}{C}/\nucl{13}{C} at least,
the observations confirm the predictions of the TNR models.

Other diatomic molecules detected in novae include CN and C$_2$
\citep[][see also references in \cite{ER-CN2}]{nagashima14,kawakita15,fujii21}.
Other than \nucl{12}{C}/\nucl{13}{C}, there have been few determinations
of isotopic ratios in nova ejecta; while \cite{pavlenko20}
present isotopic ratios for C, O and Si for the RN
T~CrB, the situation is complicated by the fact that the ratios
are for the photosphere of the red giant secondary, 
onto which the products of the TNR
on the nearby WD have been entrained.

The environment of a nova, with its intense ultra-violet radiation
field, shocks etc., is extremely hostile to small molecules and
nucleation sites, and it is very likely that molecule and grain
formation occur in dense clumps, similar to those seen in the
Helix Nebula \citep[NGC 7293;][]{hora06}. Indeed clumps with extended tails,
similar to those in the Helix, are seen in the ejected shell
of DQ~Her \citep{vaytet07}.

Detailed analysis of the 
formation of small molecules in nova ejecta has been carried out by
\cite{rawlings88}, \cite{rawlings89}, and \cite{pontefract04}.
\citeauthor{rawlings89} have followed the chemical networks as 
far as the formation of small carbon clusters, C$_n$, with $n$ 
as large as~8. \cite{derdzinski17} have shown that radiative 
shocks in nova ejecta can result in the high densities required 
for the formation of CO and of the requisite nucleation sites.

The usual paradigm is that O-rich condensates
(e.g. silicates) form in an environment in which C$<$O by number, 
while C-rich condensates (e.g. carbon, SiC) form when C$>$O. Many 
TNR models predict that 
nova ejecta are O-rich. However, \cite{jose16} explored the 
effect of the WD composition on the outcome of the TNR, 
for a WD progenitor having mass 8\Msun. 
These profiles lead to C-rich (rather than O-rich) ejecta. 
Whether the ejecta are O- or C-rich seems to depend critically on 
the composition of the outer WD shell.
Notwithstanding these issues, the above dust condensation paradigm
does not apply in novae as it is
common to find that dust-forming novae form both types of condensate
in the course of the same eruption 
\citep[see e.g.][and references therein]{ER-CN2}. Examples of this
behaviour include V705~Cas \citep{evans97,evans05,gehrz95}, 
V842 Cen \citep{hyland89, smith94}, QV~Vul, V1065~Cen \citep{helton10}, 
and V5668~Sgr \citep{gehrz18}. 
Of the these, V842 Cen, QV Vul, and V705 Cas had
deep minima in their visual light curves whereas V1065~Cen 
had a more modest dip. 

The significance of 
the C:O ratio is that, whichever of C and O has the lower abundance
is locked up in the CO molecule, which is very stable at the 
temperatures ($\ltsimeq2000$~K) at which grains form. This may 
indicate that CO formation does not go to saturation, so that both C
and O are available for incorporation into dust, or that there are steep 
abundance gradients in the ejecta, such that the C:O ratio varies
as a function of location.

\begin{landscape}
 
\begin{table*}
\centering
\caption{Indicative values of dust masses in nova ejecta.
A bullet ($\bullet$) indicates that the component was present 
but there is no mass estimate.\label{dust-mass}}
 \begin{tabular}{lcccccl}
  \multicolumn{1}{c}{Nova} & Year of  & \multicolumn{3}{c}{Dust mass ($10^{-8}$\Msun)} & Aromatic IR & \\ \cline{3-5}
  & outburst & Carbon   & Silicate & Total  & features & \multicolumn{1}{c}{Reference}\\   \hline
  NQ Vul     & 1976 & 35  &   &  & & \cite{ney78} \\
  LW Ser     & 1978 & 160 &   &  & & \cite{gehrz80a} \\
  V1668~Cyg  & 1978 &  &&  2.5 & & \cite{gehrz80b} \\
  V1370 Aql  & 1982 &    & $\bullet$ &  40--2000 &  & \cite{snijders87} \\  
             &      &    &            &            &  & \cite{bode84} \\
  QU Vul     & 1984 &     & 1   &  & & \cite{gehrz86} \\
  QV Vul     & 1984 &     &     & 100 & $\bullet$ & \cite{gehrz92}\\
  V842 Cen   & 1986 &  5   &  20    &    & $\bullet$ & \cite{hyland89} \\
             &       &    &         & &  & \cite{smith94}\\
  V838 Her   & 1991 &     & 0.48 & & $\bullet$ & \cite{woodward92} \\
  V705~Cas   & 1993 &  60--82 & 7 && $\bullet$ & \cite{gehrz95}\\
             &      &         &    && & \cite{mason98} \\
  V1424 Aql  & 1995 &         &    &  17--80 & & \cite{mason96}\\
  V5668~Sgr  & 2005 &  &&  12 & & \cite{gehrz18} \\
  V2361~Cyg  & 2005 &  &&     & $\bullet$& \cite{helton11} \\  
  V2362~Cyg  & 2006 &  & &     & $\bullet$ & \cite{helton11} \\
  V1065~Cen  & 2007 &  & $\bullet$ &  2--37 & & \cite{helton10} \\ 
  V1280~Sco  & 2007 &   6.6--8.7   & 34--43  & &$\bullet$ & \cite{sakon16} \\
  V339 Del   & 2015 & 12   &  & & & \cite{gehrz15} \\
             &      &  2.7    &  & &  & \cite{evans17} \\
  V5668 Sgr  & 2015 &   12   &$\bullet$  &  &  & \cite{gehrz18} \\  \hline
 \end{tabular}
\end{table*}

\end{landscape}

\begin{figure*}
\centering
  \includegraphics[width=1.0\textwidth,keepaspectratio]{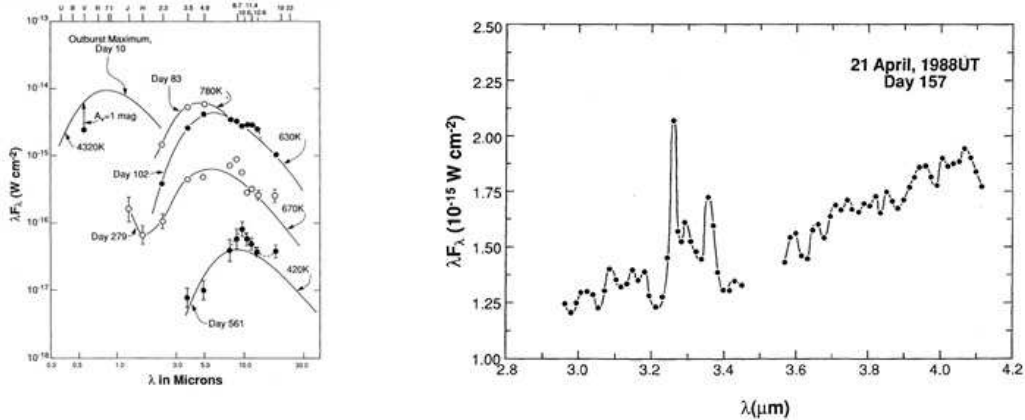}
 \caption{Evidence for the formation of four different mineral 
 compositions in nova QV~Vul. Left panel: Spectral energy distributions 
 indicating the presence of carbon, SiC and silicates at various 
 phases of the temporal development of the nova.
 Right panel: The 3.2\mic\ emission feature indicates the presence 
 of hydrocarbons.  Figures from \cite{gehrz92}. 
 \copyright{AAS}. Reproduced with permission.
\label{qvvul}}
\end{figure*}

{An extreme example of this behaviour was seen
in QV~Vul \citep[see Fig.~\ref{qvvul};][]{gehrz92}. On the evidence of 
the visual light curve, dust formed in the ejecta of this nova
at $\sim50$~days after the eruption. Carbon dust was first identified
at 83~days, and was present throughout the 2~years of observation. 
SiC (11.3\mic) appeared at 102~days, and the 9.7 and 19.5\mic\ silicate features on day 561.}

The mass of dust produced may be estimated if (a)~the emissivity
of the dust and (b)~the distance to the nova are known. Both of
these quantities can be subject to large uncertainties,
although the situation with regard to distances has improved
considerably with data from the {\it Gaia} mission
\citep{schaefer18,gaia}.
Nevertheless the dust masses are plausibly consistent with
the deduced masses of ejected gas and a reasonable gas-to-dust ratio.
A selection of dust masses determined for the ejecta of several novae
is given in Table~\ref{dust-mass}. It should be borne in mind that 
these masses were determined in various ways, from various data-sets, 
and some give total dust mass, whatever the dust composition. 
Also, detailed information about dust mineralogy became 
available only after the availability of data at reasonably
high spectral resolution; earlier studies therefore provide
little or no information about the nature of the dust. 
The data in Table~\ref{dust-mass} are given here only to 
give an indication of the range of dust masses and compositions,
and are not meant to be a rigoroursly defined sample.

However, these data enable us to estimate the 
rate at which nova dust is dispersed into the interstellar medium.
Determining the Galactic nova rate is not straight-forward, and
recent studies \citep{shafter17,de21} suggest 
$\sim50$~yr$^{-1}$. If 30\% of these produce $\sim10^{-6}$\Msun\ of
dust (cf. Table~\ref{dust-mass}) then $\sim1.5\times10^{-5}$\Msun\
of nova dust, of all types, is ejected into the interstellar medium per year.

\subsection{The nature of the dust in novae}

The early IR observations of novae during their dust phase were
broad-band photometry, which gives very little information about the 
nature of the dust, other than that the spectral energy distribution
resembled a featureless black body with a temperature that declines as
the dust moves away from the site of the explosion (see Fig.~\ref{nqvul}). 
In view of the expected overabundance of carbon in nova ejecta 
{(see Section~\ref{nova-abun})}.
the  natural conclusion was that the dust consisted primarily 
of amorphous carbon. The formation of carbon dust in nova ejecta was 
first explored in detail by \cite{clayton76}.

With the advent of IR spectroscopy at moderate resolution, 
more detailed information was available about the mineralogy of 
the dust. In particular the InfraRed Spectrograph \citep[IRS;][]{houck04}
on the {\it Spitzer Space Telescope} \citep{werner04, gehrz07}
has provided a rich source of information about the mineralogy of nova dust.

\subsubsection{Silicate dust}

\begin{figure*}
\centering
 \includegraphics[width=0.9\textwidth,keepaspectratio]{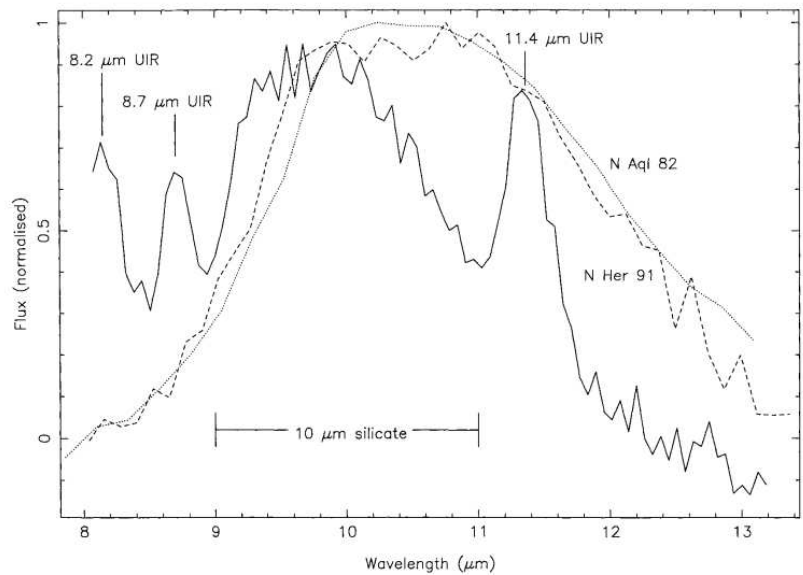}
 \caption{The 9.7\mic\ silicate features in V838~Her (broken line), 
 V1370~Aql (dotted line) and V705~Cas (solid line), from \cite{evans97}. 
 Note the ``UIR'' hydrocarbon features (referred to as AIR in this chapter)
 superimposed on the silicate feature in V705~Cas. 
  Reproduced with the permission of A. Evans.\label{Cas93_II_Sil}}
\end{figure*}

The usual signature of silicate dust is emission in the 9.7\mic\ 
Si--O stretch and 18\mic\ O--Si--O bend modes. However
these features are not good diagnostics for the precise
mineralogy of the silicate as the 9.7 and 18\mic\ features
are, by their nature, common to all silicates. Discrimination
between the various flavours of silicate (e.g. enstatite
forsterite, diopside) generally requires high 
signal-to-noise spectroscopy beyond $\sim30$\mic\ \citep{molster02},
data that are not yet available for novae.

Nevertheless there do indeed occur variations in the 9.7\mic\ silicate
feature in novae, as seen in Fig.~\ref{Cas93_II_Sil} 
\citep[from][and references therein]{evans97}, which shows the 9.7\mic\
feature in three novae: V838 Her, V705~Cas and V1370~Aql. 
It is clear that the features in V838~Her and V1370~Aql were broadly
similar, and much broader than that in V705~Cas.
\cite{smith94} attributed the breadth of, and structure in, the 
9.7\mic\ feature in novae to a degree of crystallinity,
whereas the narrowness of the feature in V705~Cas suggests that
the silicate in this nova was amorphous.
The latter also showed a very weak 18\mic\ feature, which 
\cite{evans97} attributed to the fact that the dust was freshly-formed.
This conclusion was based on the laboratory work of \cite{nuth90},
who found that the strength of the 18\mic\ feature relative to that of
the 9.7\mic\ feature strengthens as the dust is annealed.

\begin{figure*}
\centering
 \includegraphics[width=14cm]{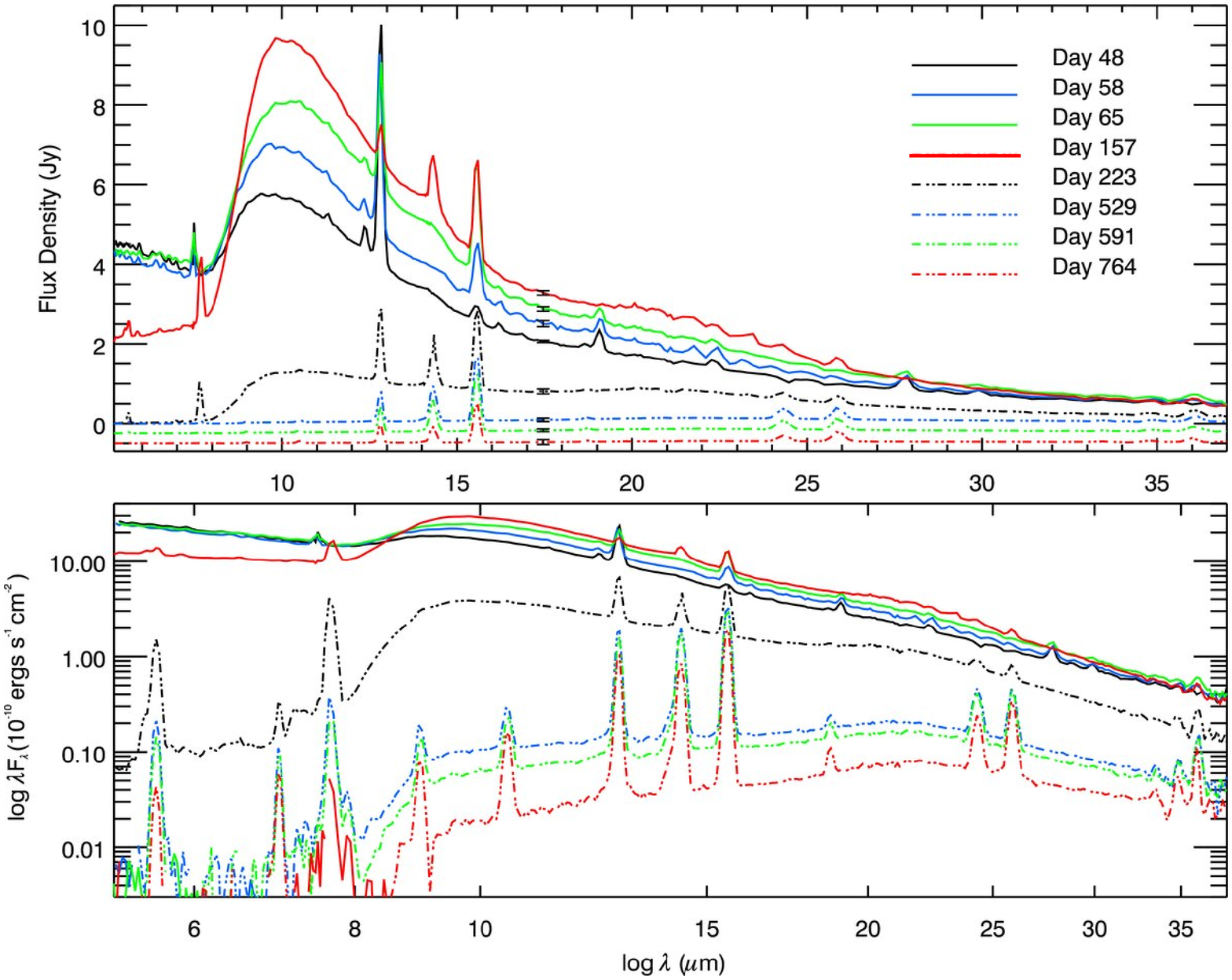}
 \caption{The evolving spectral energy distribution of the newly
 formed dust around nova V1065~Cen. Note the strong ionic emission 
 features superimposed on the dust continuum, at 7.6\mic\ (\fion{Ne}{vi}), 
 12.8\mic\ (\fion{Ne}{ii}), 13.5\mic\ (\fion{Mg}{v}), 
 15.5\mic\ (\fion{Ne}{iii}), 14.3\mic, 24.3\mic\ (\fion{Ne}{v}) 
 and 25.9\mic\ (\fion{O}{iv}). There are also hydrogen recombination
 lines at 5.7\mic, 6.3\mic, 7.5\mic, 11.3\mic, 12.4\mic, 16.2\mic.
 {Top panel shows $f_\nu$ against $\lambda$ to
 highlight the strong 9.7\mic\ silicate feature;
 the final three observations only (Days 529, 591 and 764) have been
 multiplied by a factor of 10 and offset for clarity.
 The error bars at 17.5\mic\ show representative $3\sigma$ errors.
 Bottom panel shows $\log[{\lambda{f}_\lambda}]$ against $\lambda$; no
 offsets or other factors in this panel.} From \cite{helton10}.
 \copyright{AAS}. Reproduced with permission.
\label{dust_seds}}
\end{figure*}

Fig.~\ref{dust_seds} shows the {\it Spitzer} IRS spectrum of nova
V1065~Cen, which displays strong evidence for silicate dust, 
superposed on a continuum that appears to arise from amorphous carbon.
However, analyis by \cite{helton10} of the dust emission using the
{\sc dusty} code \citep{dusty} suggested that the dust was 
dominated by silicate, with little if any contribution from 
amorphous carbon. In particular there were no hydrocarbon emission
features (see below) in this nova.  The prominence of neon lines in this 
nova indicate that it was a neon nova that originated on an ONe 
WD\footnote{The prominence of neon lines is a necessary, but not
sufficient, condition, that a nova be classified as a ``neon
nova'', which requires a neon overabundance relative to
solar of $\gtsimeq10$ \citep{helton12}.}.

\begin{landscape}
 
\begin{table*}
\centering
\caption{Primary AIR features, and their wavelengths in the ejecta of 
dusty novae. \label{air}}
 \begin{tabular}{ccccccc}
$\lambda$ ($\mu$m) & Assignment$^*$ & V842 Cen$^{{\dag},a,b}$ & V705 Cas$^c$ & DZ Cru$^d$ & V2361 Cyg$^e$ & V2362 Cyg$^{e,f}$ \\\hline
\multicolumn{7}{l}{~~}\\
3.28 & Aromatic C\chemone{H} stretch & 3.28&  3.28 & --- & --- & ---  \\
3.4  & Aliphatic C\chemone{H} stretch & 3.4 &  3.41 & ---  & ---  & --- \\
6.2  & Aromatic C\chemone{C} stretch  & --- & --- & 6.46  & 6.37--6.41 & 6.31--6.36\\
7.7--8.2 & Aromatic C\chemone{C} stretching bands& 7.7&8.06-8.17& 8.12& 7.96--8.12& 7.77--7.90\\
8.7  & Aromatic C\chemone{H} in-plane bend &8.6 & 8.7 & $\circ$ & $\circ$ & $\circ$ \\
9.2  &                        & $\circ$ & 9.29 & 9.26 & $\circ$ & 9.52--9.83 \\
11.5 & Aromatic C\chemone{H} out-of-plane bend & 11.3 &11.40 & 10.97 & 11.37--11.50 & 11.25--11.40 \\ 
12.7 & C\chemone{H} out-of-plane bend & 12.7 & $\circ$ & 12.36 & 12.33--12.40 & 12.46--12.50\\
&&&&&&\\\hline
\multicolumn{7}{l}{~~}\\
\multicolumn{7}{l}{$^*$A ``$\circ$'' indicates that the relevant feature
was weak or absent, a ``---'' that the observations did not cover the}\\ 
\multicolumn{7}{l}{relevant wavelength range.}\\
\multicolumn{7}{l}{$\dag$The wavelengths of the features in papers (a) and (b) were simply reported as the ``family of emission features}\\
\multicolumn{7}{l}{at 3.28, 3.5, 6.2, 7.7, 8.6 and 11.3\mic''.}\\
\multicolumn{7}{l}{(a): \cite{hyland89}; (b): \cite{smith94};
(c): \cite{evans05}; (d): \cite{evans10};}\\
\multicolumn{7}{l}{(e): \cite{helton11,helton14}; (f) \cite{lynch08}.}\\
 \end{tabular}
\end{table*}
\end{landscape}

\subsubsection{Hydrocarbons\label{nova-air}}

\begin{figure*}
\centering
 \includegraphics[width=1.00\textwidth,keepaspectratio]{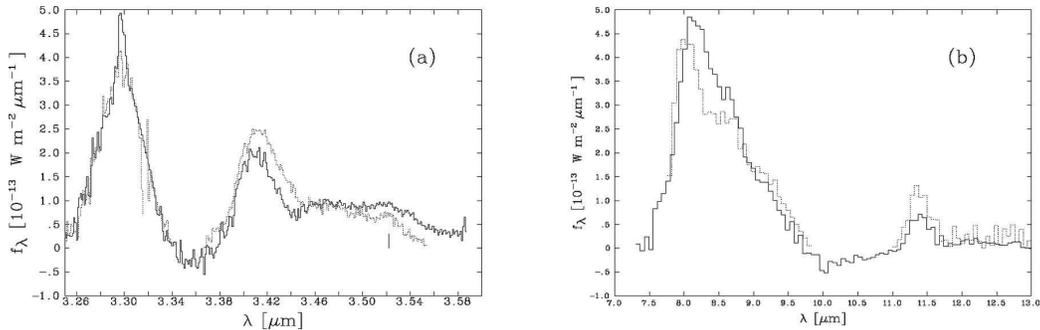}
 \caption{The AIR features in nova V705~Cas.
 Left: 3\mic\ band. Note the strength of the 3.4\mic\ feature relative to that of
 the 3.28\mic\ feature.
 Right: 6--12\mic\ band.  
 {In each case the solid line denotes data from 1994
 August (day~253), the dotted line from 1994 October/November (day~320).}
 From \cite{evans05}.
 Reproduced with the permission of A. Evans.\label{Cas93_IV_UIR}}
\end{figure*}

\begin{figure*}
\centering
 \includegraphics[width=1.0\textwidth]{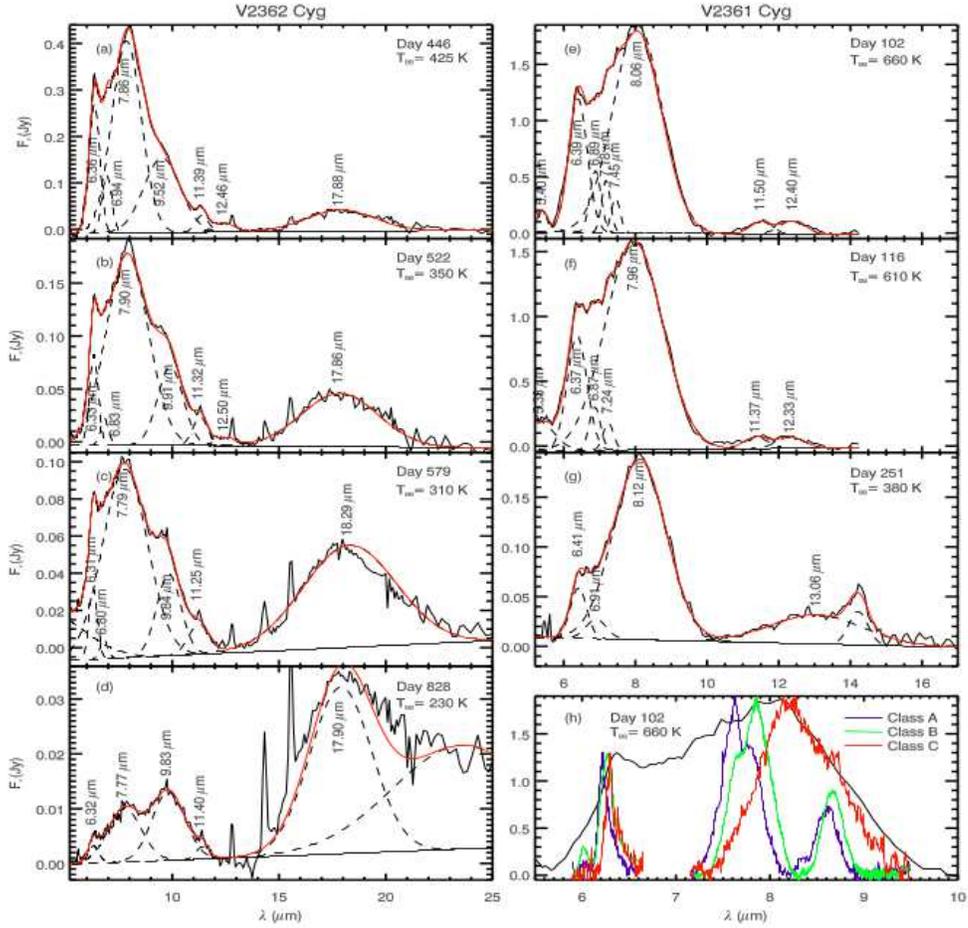}
 \caption{AIR emission in novae V2362~Cyg (left) and V2361~Cyg (right).
 In both cases a blackbody continuum has been subtracted to highlight
 the AIR features. 
 {In panels (a)--(g) the dashed lines are individual AIR 
 features, fitted with Gaussian components to produce the composite fit 
 (red); peak wavelengths of individual AIR features are indicated
 (compare with Table~\ref{air}).
 Panel (h) shows Class A, B, and C UIR profiles \citep{peeters02} 
 overlayed on the V2361 Cyg spectra from day 102.} 
 In V2362~Cyg the
 features around 9.8\mic\ and 18\mic\ are most probably due to silicates.
 From \cite{helton11}.  \copyright{AAS}. Reproduced with permission.
\label{UIRs}}
\end{figure*}

The possibility that hydrocarbon features are present (either from
free-flying Polycyclic Aromatic Hydrocarbon (PAH) molecules,
or from Hydrogenated Amorphous Carbon (HAC)) was predicted
by \cite{mitchell84}. These features (to which we shall refer
as Aromatic Infrared (AIR) features) have now been detected
in several novae, courtesy of their emission at 3.3\mic, 3.4\mic,
6.6\mic, 7.7\mic\ etc. A selection is shown in Figs.~\ref{qvvul},
\ref{Cas93_IV_UIR} and~\ref{UIRs}.
Note how, in V2362~Cyg, the AIR feature wavelengths and relative
fluxes change as the ejecta evolve.

In the majority of astrophysical sources in which they are observed, 
the main AIR features are seen at 
\citep[see][for comprehensive compilations]{allamandola89,tielens08}
3.29\mic, 3.4\mic, 6.2\mic, 7.6--8.0\mic, 8.7\mic\ and 11.2\mic,
which are attributed to various bend, stretch  and wag modes in large
PAH molecules \citep{tielens08}.

A list of the more prominent AIR features, and their wavelengths as
measured in novae, is given in Table~\ref{air}, which also gives 
the assignments. Note that the 9.2\mic\ 
feature is not normally recognised as part of the usual ``family'' of AIR 
features \citep[see][]{tielens08}, but \cite{ootsubo20} (see below) 
have suggested that it might be attributed to aliphatic hydrocarbons. 
Minor AIR features, also present in nova spectra, are not listed, 
but they may appear as ``shoulders'' on the main features.

The AIR features in novae seem to be unlike those seen in other
astrophysical sources, such as planetary nebulae, 
post-Asymptotic Giant Branch stars, star-forming regions
\citep[see][for a full classification of AIR features in the 6--9\mic\
wavelength range]{peeters02}. In novae the 3.4\mic/3.3\mic\ 
flux ratio is large compared with other sources. This may 
reflect the relative amounts of aromatic and aliphatic 
hydrocarbons in nova dust, or may indicate the presence of 
CH$_2$ and CH$_3$ groups in silicate grains; laboratory studies
\citep{grishko02a} have shown that such contaminants enhance 
the 3.4\mic\ AIR feature. The lower right panel of Fig.~\ref{UIRs}
compares nova AIR features with those of \citeauthor{peeters02}'s
``Class~C'', which \cite{sloan07} have attributed to a carrier having
predominantly aliphatic bonds.

The AIR feature normally at 7.7\mic\
appears at 8.2\mic\ in novae; there are also ``shoulder'' features
(at 8.7\mic\ and 9.2\mic) on the 8.2\mic\ feature that may the result 
of the incorporation of nitrogen-bearing groups into the PAH 
molecule \citep{evans05}. 
Support for this suggestion comes from
the laboratory study by \cite{endo21}, who found that
``Quenched Nitrogen-included Carbonaceous 
Composite'' (QNCC) reproduces the nova-specific AIR very well. 
The fact that the nitro\-genated PAH molecules 1- and 2-cyanonaphthalene 
are present in the nearby Taurus Molecular Cloud TMC-1 
\citep{mcguire21} hints that such compounds must surely have been
present in the nascent Solar System.

Furthermore, the 11.25\mic\ feature in novae appears 
at longer wavelength (11.4\mic). As the \nucl{12}{C}/\nucl{13}{C}
ratio is low in novae, some of these differences may be attributed 
to isotopic shifts \citep[see][]{wada03}. However the wavelength
shifts are more likely due to the incorporation of other species
such as N into the PAH structure.

\subsection{Summary}

Classical nova explosions are laboratories in which several 
important, but poorly understood, astrophysical processes can 
be observed in real time; in particular they provide the 
opportunity to observe dust formation. Not only do they produce 
a variety of grain types, they often form both oxygen- and 
carbon-rich condensates simultaneously. In principle it is possible to 
see gas-phase abundances decrease as dust forms and atoms are removed
from the gas, although this has been directly observed 
only in the case of V1370~Aql \citep{snijders87}.

\section{``Born-again'' giants}

\subsection{The Very Late Thermal Pulse}

It is well known that, when a star evolves away from the 
Main Sequence  (MS), its post-MS evolution depends on its mass. 
For stars having low to intermediate mass 
{\citep[$\ltsimeq8$\Msun;][]{werner06}}, following the helium
core flash, burnout of He occurs in the core on the horizontal 
branch. It then evolves up the AGB, 
and the star sheds its outer envelope which remains visible as 
a planetary nebula (PN), irradiated by the still-hot stellar core. 
In time the PN disperses and the stellar core becomes a WD.

However, in as many as 10--20\% of cases \citep{blocker01,lawlor03},
after the star has evolved beyond the post-AGB phase and is evolving 
towards the WD region, it re-ignites a residual helium shell in a VLTP. 
It then retraces a large part of its evolutionary track across the 
Hertzsprung-Russell diagram \citep{iben82,herwig01,lawlor03} 
and becomes a ``Born-Again Giant'' (BAG). The final evolution,
from pre-WD to the BAG phase, was predicted to 
take of the order of a few centuries, thus representing a very rapid 
(and hence seldom seen) phase of stellar evolution. 

Whereas several hundred classical novae have been catalogued and studied, 
both in the Galaxy and in galaxies as far as the Virgo cluster, VLTPs are 
very rare indeed. Their number is, as of the time of this writing, fewer
than ten. These include Sakurai's Object (V4334~Sgr), V605~Aql 
\citep{clayton13} and FG~Sge \citep{gehrz05}, although the latter has 
also been regarded as a Late Thermal Pulse\footnote{See \cite{werner06}
for the difference between a Late Thermal Pulse and a
{\it Very} Late Thermal Pulse.} \citep{jeffery06}. Most of 
these objects are hydrogen-deficient\footnote{However \cite{blocker97} 
note that FG~Sge is not hydrogen deficient.}, carbon-rich, 
have extensive dust shells, and each lies at the centre of an old PN. 
The post-VLTP evolution of Sakurai's Object, FG~Sge and V705~Aql have 
been very similar \citep{hinkle08,clayton06,clayton13} and there is 
a suggestion that they represent different stages along 
the same evolutionary sequence \citep{lawlor03}.

While it is now understood that they have their origins on 
the pre-WD core of an AGB star, whether these are isolated objects, 
or in binary systems, is unclear; the dust disks around V605~Aql 
\citep{clayton13} and Sakurai's Object \citep{chesneau09} 
may hint at binarity. 
Indeed, an
alternative interpretation
of the phenomenon typified by Sakurai's Object is the rapid accretion
of material onto a WD, presumably from a companion 
\citep{denissenkov17}. In such a scenario the flash is recurrent
(just as in the case of a nova) and if correct, means that
individual objects might make a more substantial contribution to the
interstellar medium than is the case for a single, one-off,
VLTP.

Detailed modelling by \cite{herwig11} and others,
based mainly on the behaviour of the well-studied Sakurai's Object,
shows that the overall features of VLTPs, such as abundance and
isotopic anomalies, can be understood in terms of
ingestion of H into the He-shell convection zone. In a short 
time \citep[typically $\sim$~an hour;][]{herwig11}, there occurs 
a separation of the convection zone into the original zone
driven by He-burning, and a new zone driven by the rapid burning 
of the ingested H. While the process does not have the explosive 
nature of a nova TNR, it does nevertheless occur on an extremely short 
time-scale.

\begin{figure*}
 \includegraphics[width=0.5\textwidth,keepaspectratio]{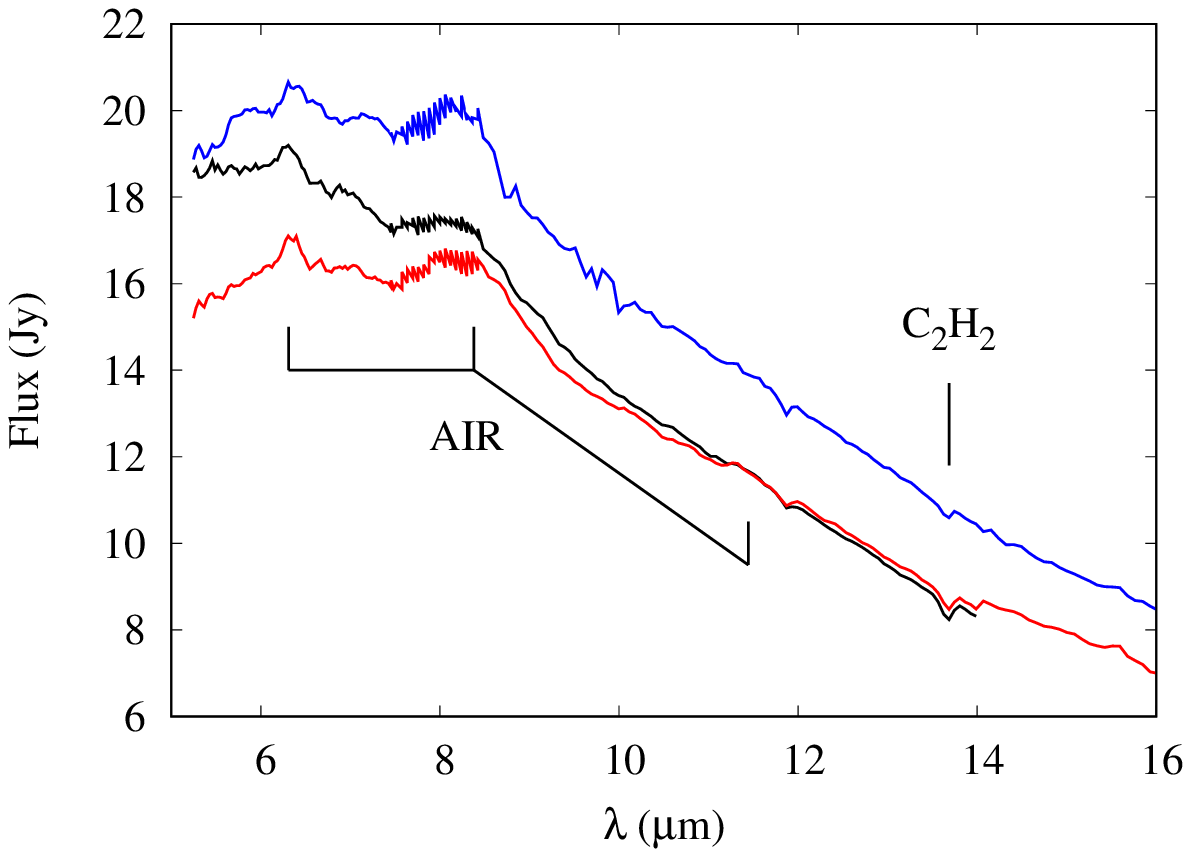}
 \includegraphics[width=0.5\textwidth,keepaspectratio]{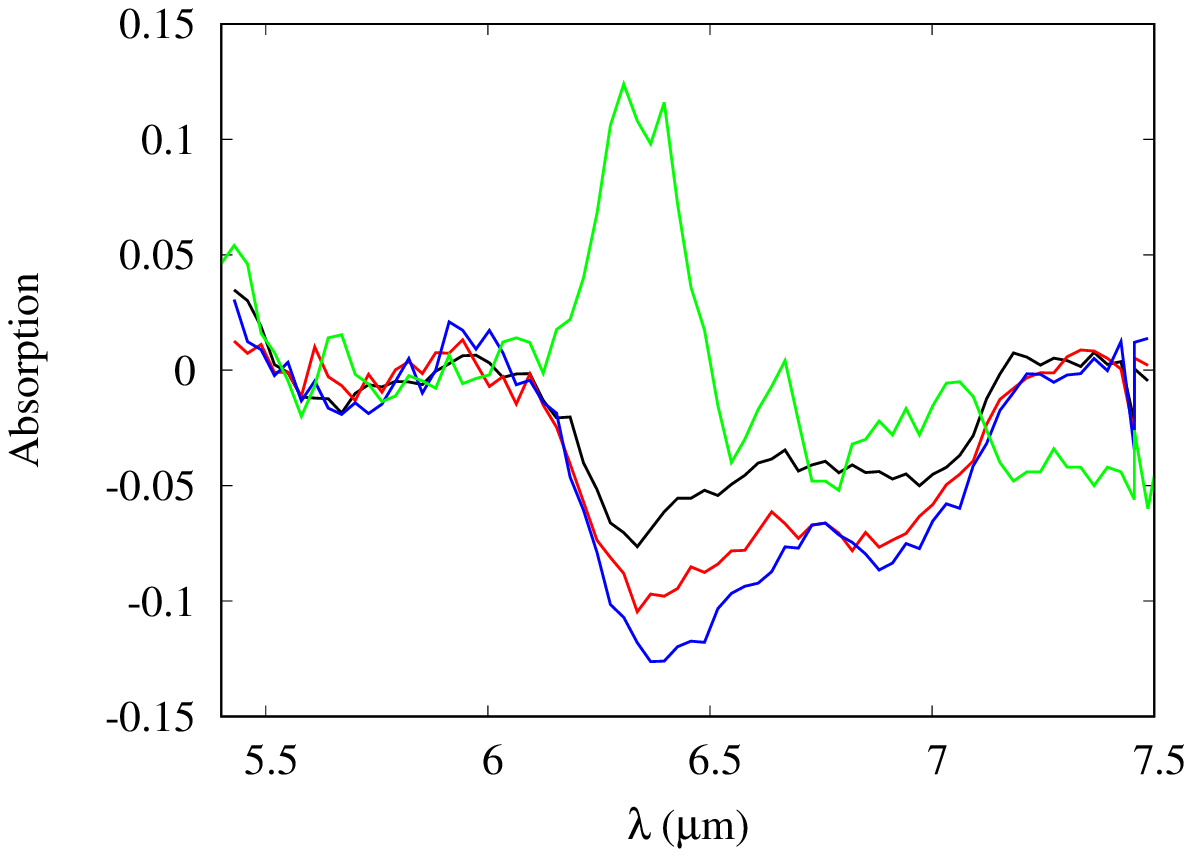}
 \caption{Left panel: Spitzer spectra of FG~Sge, showing AIR features and 
 C$_2$H$_2$. Right panel: 6.4\mic\ absorption features in Sakurai's Object,
 rectified by quadratic continuum (red, blue and black curves denote observations
 obtained on different dates); green curve is 6.4\mic\ AIR feature in FG~Sge.
 Evans et al. (in preparation).
 \label{6mic}}
\end{figure*}

Elemental abundances in Sakurai's Object were 
determined shortly after its VLTP by \cite{asplund99}.
These authors found that, in 1996, it was H-deficient, slightly
metal-poor by comparison with solar values, and
had a high abundance of some (but not all) $s$-process elements
such as Cu, Zr and Y.
They concluded that Sakurai's Object had undergone 
H-burning, followed by He-burning, a second phase 
of CNO-cycling, and $s$-processing
for light s-process elements.
The low \nucl{12}{C}/\nucl{13}{C} ratio
\citep[$\simeq6$; see e.g.,][]{pavlenko04,worters09,evans06,evans20}
also points to second stage CNO-cycling following He-burning.
The scheme described by \cite{herwig11} results in an
over-production of Rb, Sr, and Y 
\citep[over-abundant relative to solar by $\sim1.5-2$~dex 
in Sakurai's Object;][]{asplund99} that is two orders of magnitude 
higher than that of Ba and La (under-abundant relative to solar
in Sakurai's Object).

In late 1997 Sakurai's Object ejected a carbon
dust shell that completely obscured the central star. At the time of
this writing the star remains obscured.
The IR SED of Sakurai's Object is almost featureless, and closely 
resembles a cooling black body with absorption features superimposed
\citep{evans20}. Assuming that the dust is amorphous
carbon, \cite{evans20} estimated that the mass of dust resulting from
the 1997 ejection event is $\sim2\times10^{-5}$\Msun, although
dust ejection did not end with the 1997 event 
\citep{evans20,evans22}. 
A similar dust mass ($\sim7\times10^{-6}$\Msun) has been determined 
for V605~Aql \citep{clayton13}.

If these dust masses are representative
(based of course on {\it very} small statistics, and see
below for FG~Sge), then we can make a very crude extimate of
the rate at which VLTPs inject dust into the interstellar medium.
The formation rates for planetary nebulae 
($\sim4.4\times10^{-3}$~kpc$^{-3}$~yr$^{-1}$)
and white dwarfs 
($\sim2\times10^{-3}$~kpc$^{-3}$~yr$^{-1}$)
are very similar \citep{phillips84}.
Taking $3\times10^{-3}$~kpc$^{-3}$~yr$^{-1}$,
and assuming that 15\% of these stellar deaths undergo
a VLTP, ejecting $10^{-5}$\Msun\ of amorphous carbon
dust, then $\sim4.5\times10^{-9}$\Msun\ of carbon dust
is injected kpc$^{-3}$~yr$^{-1}$.

The smooth dust SED of Sakurai's Object
is in distinct contrast to the IR SED of FG~Sge, which
shows the AIR features at 6.4\mic, 8\mic\ and (weakly) 11.5\mic\
(see Fig.~\ref{6mic}).
Although Sakurai's Object has (as of the time of this writing) 
shown no evidence of AIR emission, \cite{evans20} noted weak 
absorption features around 6--7\mic, which they attributed to 
nitrogenated HAC. Similar features are often seen in lines
of sight through molecular clouds and environments containing 
young stellar objects \citep{boogert08}.
The temperature of the dust shell around FG~Sge
is roughly constant in time 
\citep[$\sim800-1000$~K;][]{gehrz05},
and the dust mass is $\sim6.8\times10^{-9}$\Msun.

The 6.4\mic\ feature profiles in Sakurai's Object, which in 
\citeauthor{evans20} had been obtained by rectifying with respect 
to a black body, were rather ill-defined. Rectifying with respect 
to the adjacent continuum brings the features out rather better 
(see Fig.~\ref{6mic}); these absorption features in Sakurai's 
Object seem to correspond well with the AIR emission features 
in FG~Sge (see Fig.~\ref{6mic}). A detailed analysis of these
features by \cite{bowey21} identifies them with a mix of PAHs, 
melilite, and large ($\sim20$\mic) SiC grains. Interestingly,
melilite is a crystalline silicate that is a significant
component of calcium-aluminium-rich inclusions in primitive meteorites
and \cite{bowey21} suggests that the 
melilite in the winds of objects like Sakurai's Object 
may be a precursor to meteoritic melilite. However it is 
likely that the melilite in calcium-aluminium-rich 
inclusions condensed in the early Solar System.

Laboratory work by \cite{grishko02b}, in which they carried out 
laser ablation of graphite in H$_2$/N$_2$ and NH$_3$ gas mixtures,
showed that the resulting material showed features in the 6.0--6.5\mic\ 
range. Given that \cite{asplund99} had found that, before it was 
enshrouded by the dust, nitrogen was overabundant in Sakurai's Object,
nitrogenation of the AIR carrier may be at work in Sakurai's Object 
and FG~Sge, as it might have been in the case of novae
(see Section~\ref{nova-air} above). 

\subsection{Summary}

What is known about dust in VLTPs relies heavily on observations
of Sakurai's Object and, to a lesser extent FG~Sge and V605~Aql.
Nevertheless enough information has been garnered over the past few
years to enable the possible identification of VLTP dust in material
that may have been ingested into the early Solar System.

\section{The pre-solar connection}

Stardust in the primitive solar nebula
suffered a large amount of destructive processing since the 
individual dust grains migrated from their various stellar origins, 
via the formation of the Sun and planets, to the present day. The 
result is that precious few of the ``pristine'' stellar grains remain.
However, despite the fact that recovered material is usually a 
heterogeneous mix of grains having a variety of stellar origins,
laboratory analysis of cometary materials, meteorites, 
and IDPs believed to have been 
released from comet nuclei can reveal information about 
the stellar sources of their constituent dust particles.

In attempting to identify grains that originated in novae and VLTPs
in pre-solar material, we should be mindful of the potential
ambiguities. For example, did SiC form in the ejecta of novae 
or in those of Type II supernovae
\citep{nittler04a,nittler04b,liu16}? And isotopic ratios, unless 
several are available \citep[see, e.g.,][]{iliadis18}, may not 
unambiguously point to an origin in novae, VLTPs or supernovae.
Furthermore, nova and VLTP grains
in (for example) IDPs have to be separated from the overwhelming
amount of minerals that are of solar, and other stellar, origin.

\subsection{Novae}

Evidence for a nova origin for some pre-solar grains comes 
therefore from studies of pre-solar grains identified in meteorites 
and of IDPs recovered from comet tails \citep{messenger02}. 

The most decisive connection
between nova grains and pre-solar material is the close
similarity in the isotopic ratios found
in certain cometary grains and the predictions of the TNR scenario.
As noted in Section~\ref{nova-abun}, the 
TNR imprints nova-specific isotope ratios, e.g. 
\nucl{12}{C}/\nucl{13}{C}, \nucl{14}{N}/\nucl{15}{N},
\nucl{20}{Ne}/\nucl{22}{Ne} on the nova ejecta. 
\cite{bose19} compared isotope ratios for ejecta produced in CO 
nova TNR models with isotope ratios observed in thirty pre-solar grains
from meteorites and found evidence for pure CO nova ejecta
material in five of them. They suggested that these SiC grains 
formed in the  winds of CO novae. 
Furthermore, \cite{bose19} found that the elemental 
and isotopic abundances in one pre-solar grain matched the 
prediction of TNR modelling, without the need for dilution.
\cite{iliadis18} identified six pre-solar grains
having isotopic signatures that indicated a ``high plausibility'' that
they originated in a CO nova, without assuming any dilution of the 
ejecta before grain condensation.
Additional studies 
by \cite{haenecour19} and by Leitner and Hoppe (2019) confirm the 
ability to discern grains in primitive Solar System materials
that were produced in nova events. 

\begin{figure*}
\centering
  \includegraphics[width=1.0\textwidth,keepaspectratio]{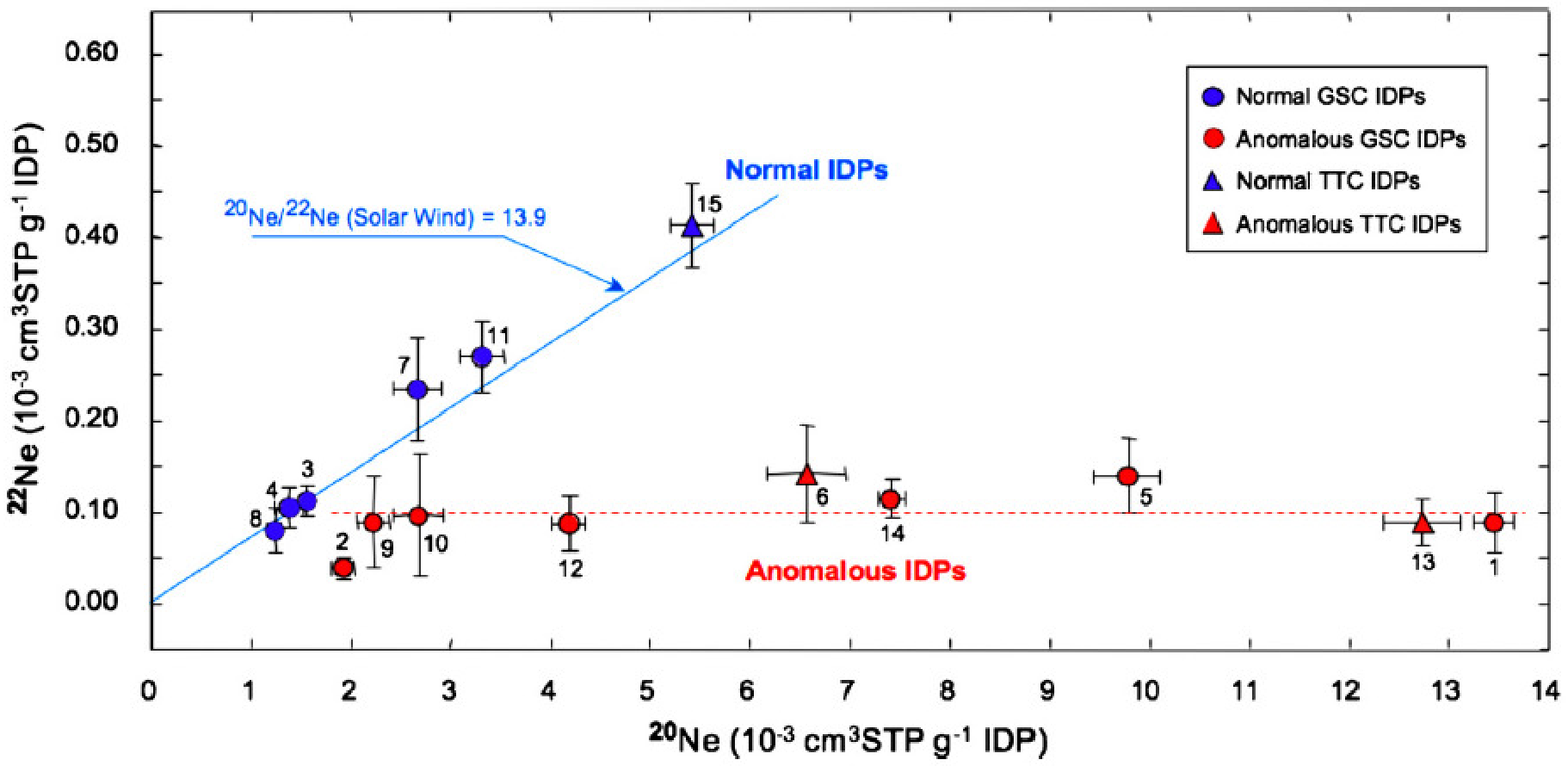}
 \caption{Concentrations of \nucl{20}{Ne} and \nucl{22}{Ne} in 15 
 IDPs from the Grigg-Skjellerup (GSC) and Tempel-Tuttle (TTC) 
 cometary collections, where 1~cm$^3$ STP contains $2.69\times10^{19}$ 
 Numbering of points corresponds to
 the IDP numbering in Table~1 of \cite{pepin11}.
 Normal IDPs show abundances expected for long-term exposure 
 to the solar wind, as would be expected for asteroidal particles
 (blue symbols and line).  
 Error bars are $1\sigma$. Details of the analysis 
 are given in \cite{pepin11}.  Figure from \cite{pepin11}.
 \copyright{AAS}. Reproduced with permission.
 \label{neon}}
\end{figure*}

Analysis of neon isotope abundances in fragmets of IDPs collected by 
a U2 aircraft from comet orbital trails suggests that TNRs from 
nova explosions on ONe WDs may also have produced some of the stellar 
debris that was present in the primitive Solar System \citep{pepin11}.
\citeauthor{pepin11} identified two distinct 
categories of IDPs having cometary origin: those (designated 
``normal'') having a \nucl{20}{Ne}/\nucl{22}{Ne} ratio (=13.9) 
and He content that is characteristic of the solar wind, and those
(designated ``anomalous'') with no detectable solar wind He and 
a \nucl{22}{Ne} abundance that is essentially uniform,
with \nucl{20}{Ne}/\nucl{22}{Ne} ratios ranging from $\sim26$ 
to $\sim153$ (see Fig.~\ref{neon}). In none of the anomalous IDPs
was He detected, clearly illustrating that 
these are a distinct population, distingiushable 
from IDPs that have had extended exposure to the solar wind.

\begin{figure*}
\centering
\includegraphics[width=0.4\textwidth,keepaspectratio]{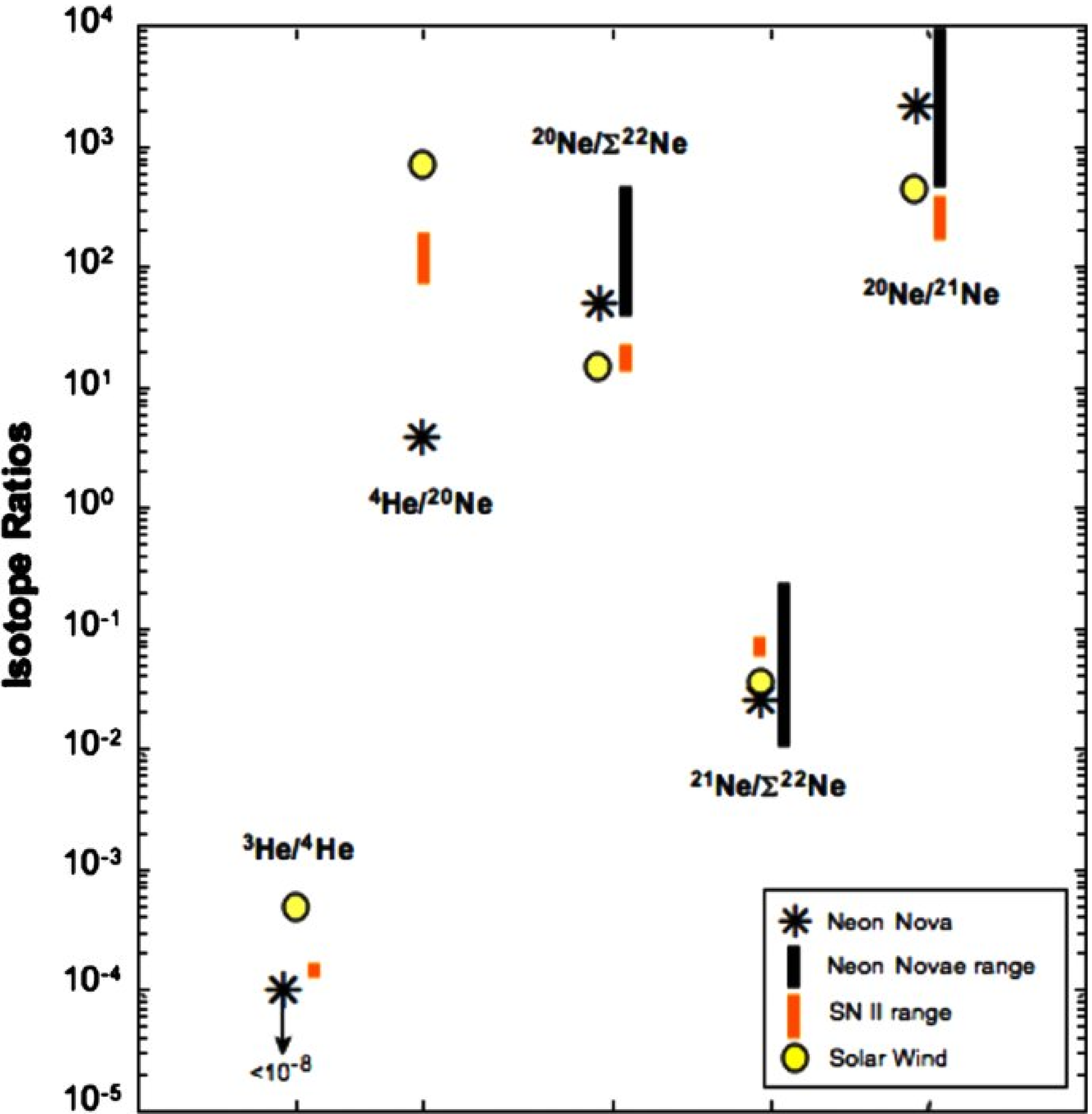}
\includegraphics[width=0.41\textwidth,clip,keepaspectratio]{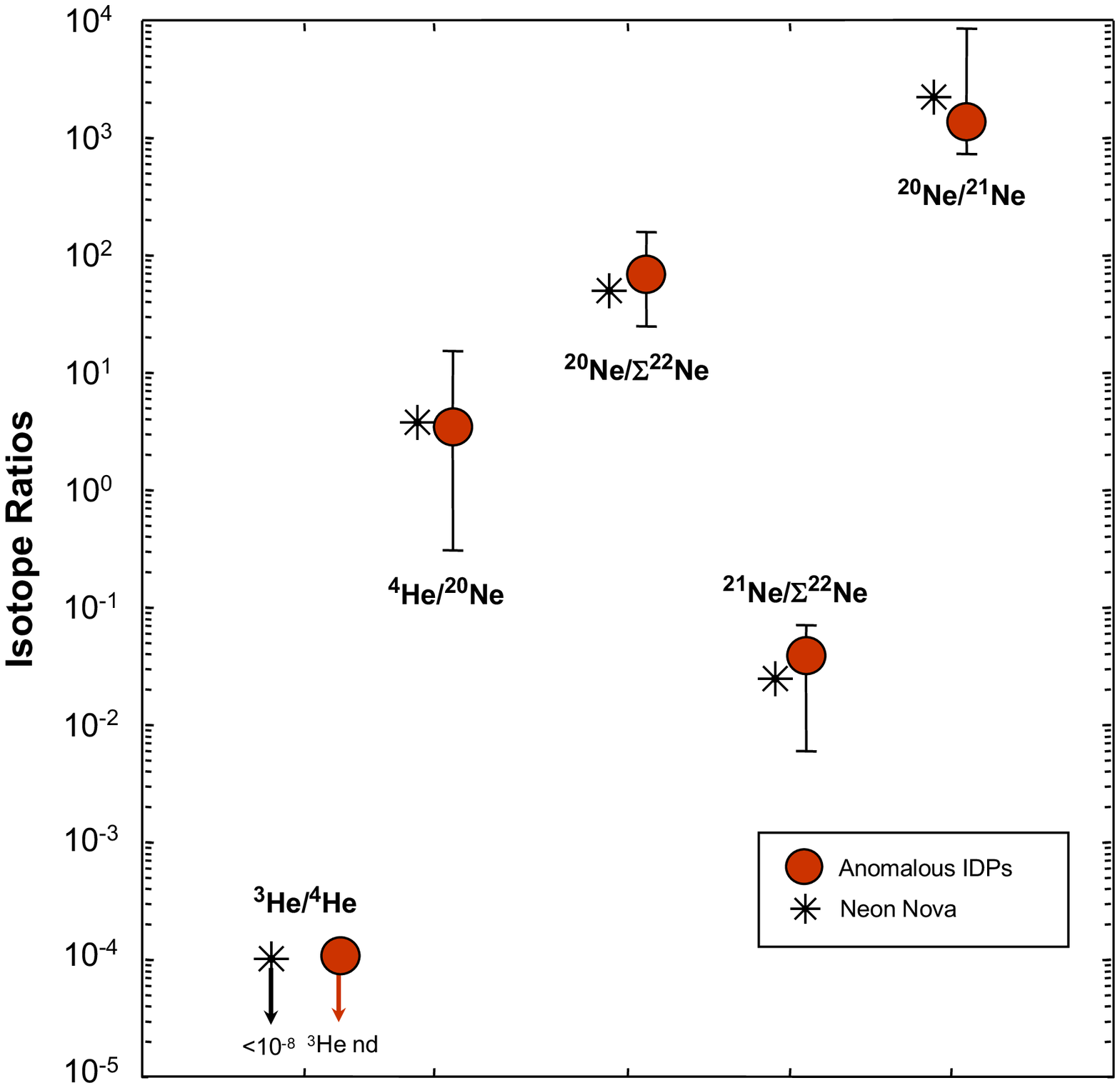}
 \caption{Left panel: Modelled He and Ne isotope ratios in neon 
 nova ejecta, with core collapse supernova models and solar wind 
 ratios plotted for comparison. The vertical black bars indicate 
 the ranges of ratios generated in models of neon nova outbursts 
 over a WD mass range of 1.00--1.35\Msun\ with several different 
 choices for initial compositions and nuclear reaction networks
 \citep{jose04}. The neon nova ejecta compositions for a 1.25\Msun\
 ONe WD are shown by the star symbols. $\Sigma^{22}$Ne represents 
 the sum of the \nucl{22}{Ne} and the $\beta^+$-active \nucl{22}{Na} 
 synthesized in the TNR. Orange bars are compositional ranges for 
 the total ejecta from models of nucleosynthesis in Type II supernovae 
 (SNe II) of 15\ and 25\Msun\ \citep[see][]{rauscher02}. Right panel: 
 Measured He and Ne isotope ratios in the anomalous 
 IDPs analyzed by \cite{pepin11} compared with solar wind 
 compositions and TNR modeling results. The IDP Data points 
 are average ratios in the anomalous groups; 
 thin vertical bars indicate
minimum and maximum measured values. ``nd'' indicates not detected.
 Neon nova modeling results for a 1.25\Msun\ ONe WD are shown by 
 the star symbols. Left panel from \cite{pepin11}; right panel
 is a modified version of a figure from \cite{pepin11} provided by
 R. O. Pepin. 
 \copyright{AAS}. Reproduced with permission.
 \label{pepin}}
\end{figure*}

By comparing the He and Ne isotopic ratios 
found in the anomalous IDPs with those predicted for 
ONe novae and core collapse
supernovae, \cite{pepin11} argued 
that several of the anomalous IDPs have He and Ne isotope ratios that
are consistent with their having been condensed 
in the outflow of a neon nova (see Fig.~\ref{pepin}).

There is some evidence that nova dust in meteorites
may have originated in the ejecta of novae that originated on both
CO and ONe WD. \cite{gyngard10} have determined isotopic ratios 
for O and Mg in grains recovered from the Murray carbonaceous chondrite
meteorite, and find that large excesses of \nucl{17}{O},
\nucl{25}{Mg} and \nucl{26}{Mg} point to an origin in a TNR on a CO
WD. \cite{starrfield97} have argued that the class of ONe novae may 
form dust grains that carry the Ne-E \citep{black69} and 
\nucl{26}{Mg} \citep{gray74} anomalies observed in meteoritic grains. 

\begin{figure*}
\centering
  \includegraphics[width=0.8\textwidth,keepaspectratio]{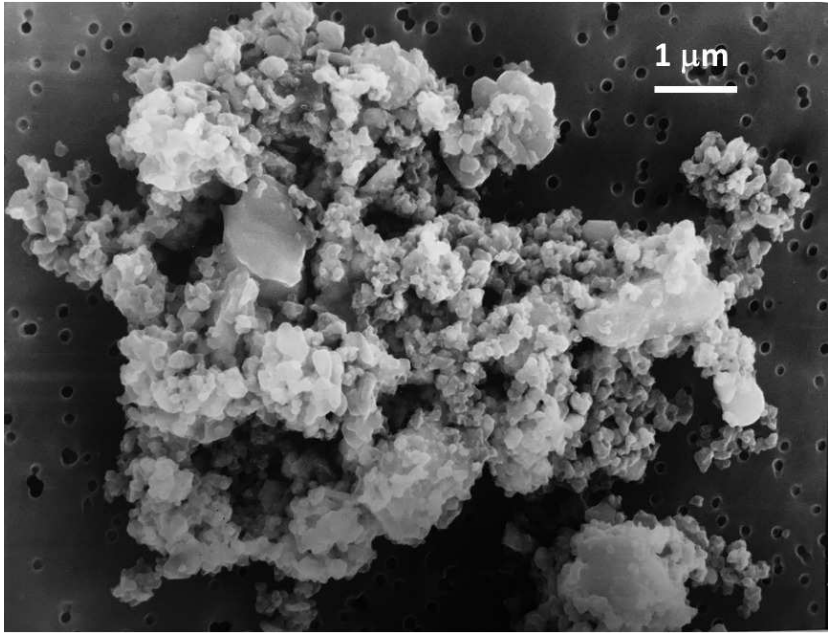}
 \caption{Electron microscope image of an 18\mic\ long IDP, possibly of 
 cometary origin, collected by a NASA high-altitude research aircraft.    
 It is a porous aggregate of unequilibrated components that were 
 never compacted by mechanical force, heating or aqueous alteration.  
 It has the elemental composition of a carbonaceous chondrite but a 
 totally different structure. This type of aggregate structure is 
 likely to be characteristic of micron and larger circumstellar 
 grains that accrete by gentle processes and not altered after formation.
 Image and caption courtesy of D. E. Brownlee. \label{brownlee}}
\end{figure*}

\begin{figure*}
\centering
  \includegraphics[width=0.8\textwidth,keepaspectratio]{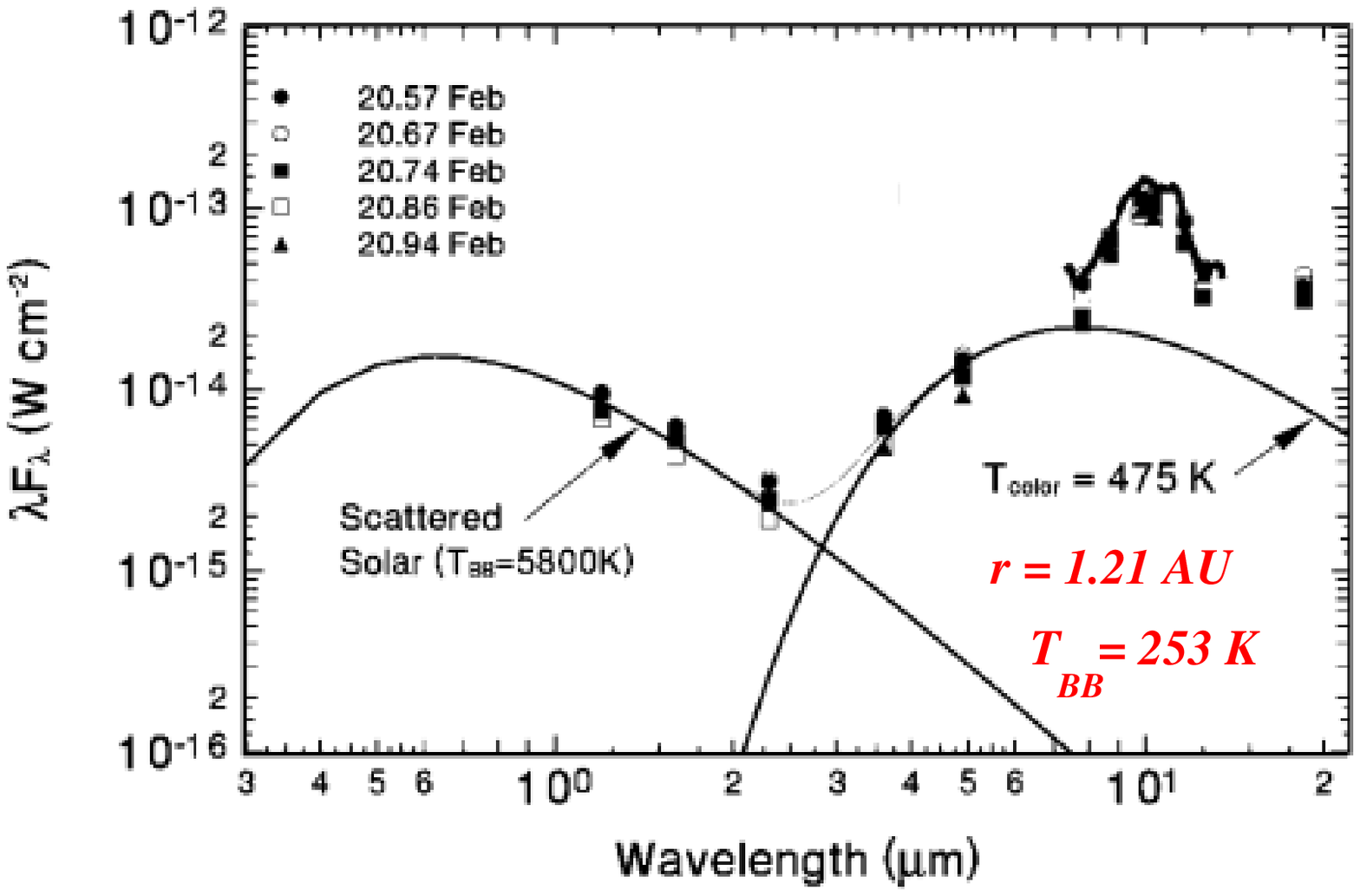}
    \includegraphics[width=0.8\textwidth,keepaspectratio]{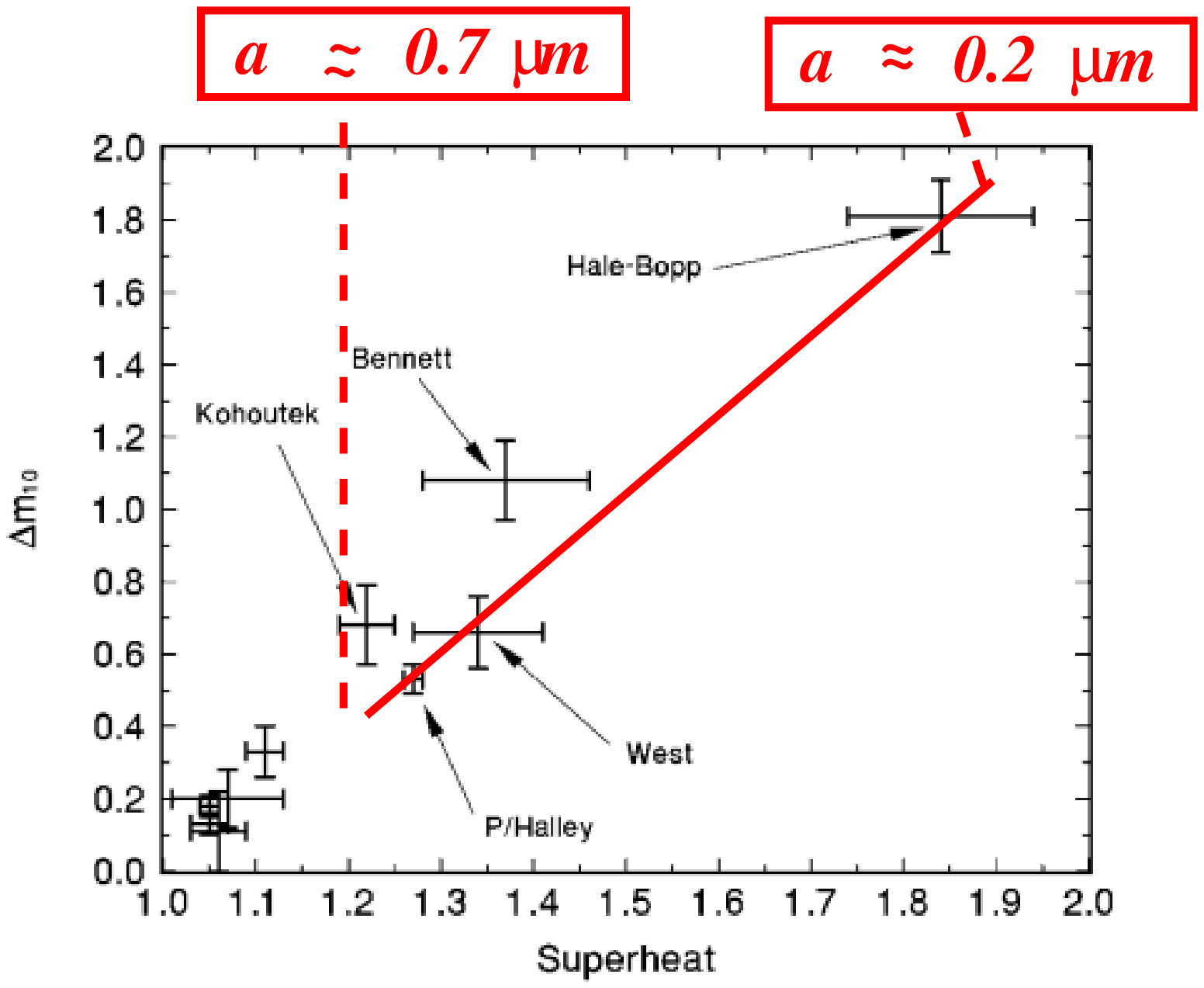}
 \caption{Top panel: Infrared SED of 
 Comet Hale–Bopp (C/1995 O1) showing a superheat (see text)
 of 1.21 and a 10\mic\ silicate feature strength of 1.8~magnitudes, 
 indicating an average grain radius of 0.2\mic. Bottom panel: Superheat 
 and 10\mic\ silicate feature strength plotted from a sample 
 of eight comets showing that comet grains have been observed 
 to span a range of 0.2\mic\ to 0.7\mic\ in radius. Nova grains 
 span the same radius range. Figure after \cite{williams97}. 
 \copyright{AAS}. Reproduced with permission.
\label{halebopp}}
\end{figure*}

Far more tentative connections between nova
dust and IDPs are in the size distribution of the particles,
and in the presence of AIR emission. The small grains in 
subclusters in some IDPs (see Fig.~\ref{brownlee}) and in
comets (see Fig.~\ref{halebopp}, in which ``superheat'' is the ratio
of the grain temperature and the temperature a blackbody would have 
at the comet's heliocentric distance; see \cite{gehrz92-com} for a 
discussion) have the same 0.1\mic\ to 0.7\mic\ size range as 
the grains that grow in nova outflows 
\citep[see, e.g.,][]{shore94,helton10,gehrz18}.
However, the size distribution is a poor discriminant,
as the presence of particles of this size are inferred in the outflows of 
evolved stars \citep[see, e.g.,][and references therein]{hofner18} and in
particular, in the ejecta of the VLTP Sakurai's Object \citep{tyne02}.

In common with a large variety of astrophysical
sources, some novae, and some comets, display AIR emission.
As noted above (Section~\ref{nova-air}), those seen in novae
\citep{evans05,evans10,helton11,helton14} differ significantly 
from those seen in other sources, possibly due to the 
incorporation of nitrogen and/or nitrogen groups into the AIR carrier
\citep{evans05}. 

The presence of AIR features in comets has been
claimed in the Infrared Space Observatory \citep{kessler96} 
spectrum of C/1995 O1 (Hale-Bopp) \citep{lisse07}, although 
these findings proved to be controversial \citep{crovisier08,lisse08}. 
The detection of AIR features in {\it Spitzer} IRS spectra of 
the debris from comet 9P/Tempel~1 following the Deep Impact 
event \citep{lisse06,lisse07} is less contentious, although these 
features resemble the ``conventional'' family of features rather 
than those seen in novae.

\begin{figure*}
 \centering
 \includegraphics[width=0.8\textwidth,keepaspectratio]{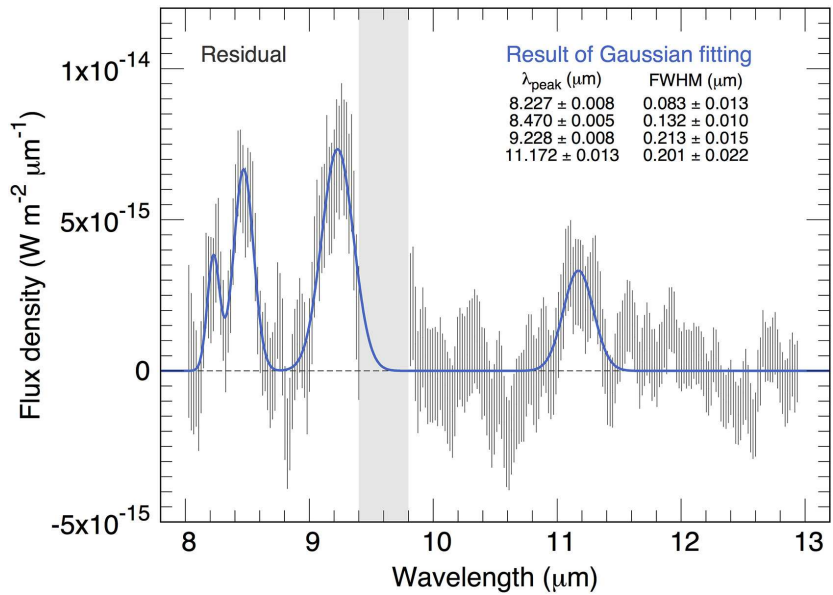}
 \caption{Possible 6--12\mic\ AIR features in comet 21P/Giacobini-Zinner.
 From \cite{ootsubo20}.  
 Figure reproduced with the permission of T. Ootsubo.\label{ootsubo}}
\end{figure*}

On the other hand, AIR features with peak wavelengths close 
to those in novae V705~Cas and DZ~Cru have been reported in 
the spectrum of comet 21P/Giacobini-Zinner \citep{ootsubo20}
(see Fig.~\ref{ootsubo}). 
The residuals in the 6--12\mic\ range, after removing the 
contibution of silicates and other components, 
revealed apparent features at 
8.236\mic, 8.475\mic, 9.229\mic\ and 11.108\mic, which 
match the AIR features in novae (see Table~\ref{air}). 
While these features were not 
detected in a {\it Spitzer} IRS observation \citep{kelley21}, 
the \citeauthor{ootsubo20} and {\it Spitzer} data were
separated in time by about 5~months. Clearly there were
differences in the modes of observation but more significantly,
there were large differences in the heliocentric distance 
(and hence cometary environment) between the two observations. 
With the exception of the silicate features,
the {\it Spitzer} IRS spectrum is feature-free
for $\lambda<14$\mic\ (M. S. P. Kelley, personal communication),
so the jury is out on the presence of AIR features in the spectrum of
21P/Giacobini-Zinner. Clearly further observations are required
to resolve this issue.

\subsection{Born-Again Giants}

Connections of VLTP dust to pre-solar grains is rather sparse by comparison 
with novae, and relies entirely on isotopic ratios for species that are
likely to have been overproduced in the VLTP. In view of the comparatively
recent (compared with novae) impetus in studing the nuclear networks in
VLTPs there is little that can be said at this stage, and much of the 
evidence is based on advancements since the VLTP of Sakurai's Object. 

\begin{figure*}
\centering
   \includegraphics[width=0.85\textwidth,keepaspectratio]{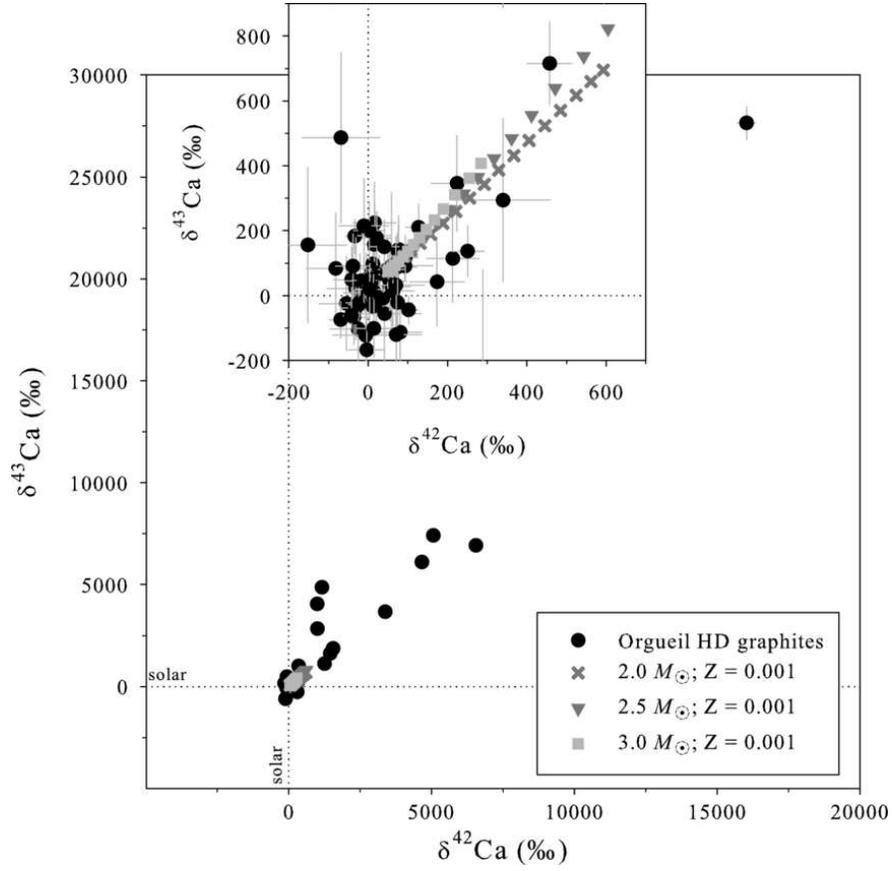}
\caption{Three-isotope plot of \nucl{42,43}{Ca}/\nucl{40}{Ca} isotopic 
ratios measured in Orgueil. Black dots are Orgueil data, the crosses, 
inverted triangles and squares are for low metallicity AGB stars.
The $\delta$ values are deviations from terrestrial ratios, 
in permil (\permil). Error bars are $1\sigma$. 
Dashed lines indicate solar ratios. Figure from \cite{jadhav13}.
\copyright{AAS}. Reproduced with permission.
\label{jadhav}}
\end{figure*}

\cite{jadhav13} have shown that the Ca and Ti isotopic anomalies in
high density pre-solar graphite grains of the 
Orgueil meteorite, combined with their low
\nucl{12}{C}/\nucl{13}{C} ratio, tally well with predictions of the 
H-ingestion phase during a VLTP event in post-AGB stars 
\cite[][]{herwig11}. The \citeauthor{herwig11} calculations were 
informed by the photospheric abundances in Sakurai's Object as 
determined by \cite{asplund99} before it disappeared from view.
In Fig.~\ref{jadhav} are plotted the ratios, as $\delta$-values, i.e.
deviations from the terrestrial ratios, in permil (\permil); this figure
also includes some predicted ratios for the envelopes of low-metallicity
AGB stars. It is evident that the high Ca isotopic anomalies measured in
Orgueil are not consistent with standard AGB models, and point to
an origin in a VLTP, which can account for the large Ca and Ti 
anomalies observed in grains with low \nucl{12}{C}/\nucl{13}{C} ratios.
\cite{amari01} proposed born-again AGB stars as one 
of the stellar sources of SiC AB grains defined as having 
\nucl{12}{C}/\nucl{13}{C} ratios $< 10$.
\cite{fujiya13} have also argued that the C, Si and S isotope 
measurements in SiC grains of Type AB are consistent with an origin
in a VLTP, but they caution that uncertainties in VLTP models mean
that a definitive conclusion is not possible.

\section{Some speculative (and provocative) concluding remarks}

Despite that fact that novae (and to a lesser extent, VLTPs) are not
expected to be major contributors to the Galactic interstellar dust 
population \citep[see, e.g., Table~1 in][]{gehrz89}, it is gratifying
that there is ample evidence that dust from these sources has survived,
and has been identified in primitive meteorites and IDPs.

Both novae and VLTPs, by virtue of their C-enriched
ejecta, contain an abundance of organic material, as evidenced by 
(novae), HCN, C$_2$H$_2$ and polyynes (Sakurai's Object), and
both types of object display AIR features. This is of course a 
trait that they share with the detritus of other low-mass stars as they
approach the stellar graveyard. 

The nova and VLTP dust would almost certainly 
have transported with them the organic material they contain.
The similarity of the organics observed in novae and in VLTPs 
(and other sources of stardust) with material present in the 
current Solar System must surely beg the question: what role might 
these organic materials have played in the origin and evolution 
of pre-biotic material in the early Solar System, and in the 
emergence of primitive biota on the early Earth?
For example, \cite{ootsubo20} (amongst many others) have suggested that 
comets aggregated and grew in the solar nebula, and that cometary 
nuclei formed in the primitive solar nebula could have delivered 
pre-biotic organic molecules to the infant Earth via impacts.

Organic material is of course widespread in the Solar System, the 
interstellar medium and beyond
\citep[see, e.g.,][for a very comprehensive review]{kwok16}. Many 
molecules detected in astrophysical environments
are precursors to the molecules that are essential for life.
For example, \cite{hadraoui19} reported the episodic detection of 
glycine (the simplest amino acid and, like the components of 
deoxyribonucleic acid (DNA), a nitrogenated/oxygenated hydrocarbon)
in comet 67P/Churyumov-Gerasimenko, while a glycine precursor has 
been detected in several molecular clouds \citep{suzuki16}. The possibility 
of detecting glycine itself in proto-planetary systems has been discussed by
\cite{jimenez14}.

As is well-known, DNA -- the cornerstone of all life on Earth -- 
consists of four nucleotides\footnote{The ``ACGT'' sequence 
known to anyone who has studied their ancestry: 
Adenine (a nitrogenated
hydrocarbon), Cytosine, Guanine and Thymine (nitrogenated/oxygenated
hydrocarbons).} linked by phosphodiester bonds. An intriguing 
consideration is that, whereas hydrocarbons are commonplace in
astrophysical environments, hydrocarbons that formed in environments
(like novae and VLTPs) in {\em which C, N, O and P are overabundant}
might be particularly relevant. The ingredients for DNA were cooked
in nova and (to a lesser extent) VLTP events.

Putting aside the issue of isotopic abundances,
do we owe our very existence to dust from novae and VLTPs that was
ingested into the proto-solar nebula?

\subsection*{Acknowledgements}

We thank Sachiko Amari, Don Brownlee, Jordi Jos\'e, 
Mike Kelley, Arumugam Mahendrasingam, Bob Pepin, Sumner Starrfield 
and Chick Woodward for their helpful comments on, and input into,
earlier versions of this chapter. 
We also thank the various authors and publishers for permission to
reproduce figures from the literature.

RDG was supported for observations presented herein by the National 
Science Foundation, NASA and the United States Air Force.

\end{document}